\documentclass[preprint,pdftex]{aastex61}

\newcommand{\chan}{{\it Chandra}}
\newcommand{\asec} {$^{\prime\prime}$}
\newcommand{\fdeg}{\mbox{$.\!^{\circ}$}}
\submitjournal{ApJ}

\shorttitle{The 4C+19.44 Jet}
\shortauthors{Harris et al.}

\newcommand{\axaf}{\mbox{\em Chandra\/}}

\usepackage[pdftex]{graphicx}

\usepackage{color}

\begin{document}


\title{A Multi-band Study of the remarkable Jet in Quasar 4C+19.44} 

\correspondingauthor{Dan Schwartz}
\email{das@cfa.harvard.edu}

\author{D. E. Harris}
\altaffiliation{Dan Harris passed away on December 6th,
    2015. His career spanned much of the history of radio and X-ray
    astronomy. His passion, insight and contributions will always be
    remembered.}
\affiliation{Smithsonian Astrophysical Observatory,
Cambridge, MA 02138, USA}

\author{N. P. Lee}
\affiliation{Smithsonian Astrophysical Observatory,
Cambridge, MA 02138, USA}

\author{D. A. Schwartz}
\affiliation{Smithsonian Astrophysical Observatory,
Cambridge, MA 02138, USA}

\author{A. Siemiginowska} 
\affiliation{Smithsonian Astrophysical Observatory,
Cambridge, MA 02138, USA} 

\author{F. Massaro}
\affiliation{Dipartimento di Fisica, Universit\`a degli Studi di 
Torino, via Pietro Giuria 1, I-10125 Torino, Italy}
\affiliation{Istituto Nazionale di Fisica Nucleare, Sezione di 
Torino, I-10125 Torino, Italy}
\affiliation{INAF--Osservatorio Astrofisico di Torino, via 
Osservatorio 20, 10025 Pino Torinese, Italy}

\author{M. Birkinshaw}
\affiliation{HH Wills Physics Laboratory, University of Bristol,
  Tyndall Avenue, Bristol BS8 1TL, UK} 

\author{D. M. Worrall}
\affiliation{HH Wills Physics Laboratory, University of Bristol,
  Tyndall Avenue, Bristol BS8 1TL, UK} 

\author{C. C. Cheung}
\affiliation{Space Science Division, Naval Research Laboratory, Washington, DC
20375-5352, USA}

\author{J. M. Gelbord}
\affiliation{Spectral Sciences, Inc.,4 Fourth Ave., Burlington, MA 01803-3304, USA}

\author{Svetlana G. Jorstad}
\affiliation{Institute for Astrophysical Research, Boston
University, 725 Commonwealth Avenue, Boston, MA 02215, USA}
\affiliation{Astronomy Department, St. Petersburg State University,
  Universitetskij Pr. 28, Petrodvorets, 198504 St. Petersburg, Russia} 

\author{Alan P. Marscher}
\affiliation{Institute for Astrophysical Research, Boston
University, 725 Commonwealth Avenue, Boston, MA 02215, USA}

\author{H. Landt}
\affiliation{Department of Physics, Centre for Extragalactic Astronomy, Durham University, South Road, Durham DH1 3LE, UK}

\author{H. Marshall}
\affiliation{MIT, Cambridge, MA 02139, USA}

\author{E. S. Perlman}
\affiliation{Physics and Space Sciences Department, Florida Institute of
  Technology, Melbourne, FL  32901, USA}

\author{L. Stawarz}
\affiliation{Astronomical Observatory, Jagiellonian University, 30-244
  Krak\'ow, Poland}

\author{Y. Uchiyama}
\affiliation{Department of Physics, Rikkyo University, 3-34-1 Nishi
  Ikebukuro, Toshima-ku, Tokyo 171-8501, Japan} 

\author{C. M. Urry}
\affiliation{Yale Center for Astronomy and Astrophysics, 260 Whitney Avenue,
  New Haven, CT 06520, USA}


\begin{abstract}

    We present arc-second-resolution data in the radio, IR, optical
and X-ray for 4C+19.44 (=PKS 1354+195), the longest and straightest
quasar jet with deep X-ray observations.  We report results from radio
images with half 
  to one arc-second angular resolution at three frequencies, plus HST and
  Spitzer data.  The \chan\ data allow us to measure
  the X-ray spectral index in 10 distinct regions along the
  18\arcsec\ jet and compare with the radio index. 
The radio and X-ray spectral indices of the jet regions are consistent
with a value of  $\alpha =0.80$ throughout the jet, to within $2\sigma$
   uncertainties. The X-ray jet structure to the south
  extends beyond the prominent radio jet and connects to the southern
  radio lobe, and there is extended X-ray emission in the direction of
  the unseen counter jet and coincident with the northern radio lobe.
This jet is remarkable since its straight appearance over a large
distance allows the geometry factors to be taken as fixed along
the jet.  Using the model of inverse Compton scattering
  of the cosmic microwave background (iC/CMB) by relativistic
  electrons, we find that the magnetic field strengths and Doppler
  factors are relatively constant along the jet. If instead the X-rays
  are synchrotron emission, they must arise from a population of
  electrons distinct from the particles producing the radio
  synchrotron spectrum.

\end{abstract}



\section{Introduction}

After more than two decades of multi-wavelength studies of
extragalactic jets, there is still no clear conclusion as to the
physical process responsible for the X-ray emission
\citep{Harris02,Harris06} from powerful quasar jets which extend to
100-kpc distances.  For low power jets, there is convincing evidence
that X-ray emission is dominated by synchrotron radiation from
electrons with Lorentz factors $\gamma$ of order 10$^7$
\citep{Worrall01,Perlman01,Hardcastle01}.  For quasar jets, inverse
Compton (iC) scattering of cosmic microwave background (CMB) photons
\citep{Tavecchio00,Celotti01} is generally invoked to model the X-ray
emission
\citep{Siemiginowska02,Sambruna02,Sambruna04,Marshall05,Sambruna06,Schwartz06,Schwartz06b,Worrall09,Marshall11,Perlman11,Massaro11}.
This mechanism requires that the energy distribution of the radio
emitting electrons extends below $\gamma\approx$~100 and that the jets
are relativistic with bulk Lorentz factors $\Gamma \approx$ 3 to 15
\citep{Schwartz15}. The bulk Lorentz factors are critical  since the
CMB energy density is enhanced in the jet rest
frame by a factor $\Gamma^2$
\citep{Dermer94,Dermer95,Ghisellini98,Ghisellini01}.  For the brighter
quasar jets, deep \chan\ observations are capable of obtaining enough
photons in spatially resolved individual regions to measure the X-ray
spectral index, $\alpha_{\rm x}$, which is one of the key parameters
in the iC/CMB model.

As noted above for low-power jets, iC/CMB may not be the only
mechanism operating. For the quasar 3C 273, both the multi-wavelength
spectra of the knots \citep{Jester06} and upper limits to \emph{Fermi}
$\gamma$-ray emission \citep{Meyer14} show that the jet must    have 
 an additional component of
radiation, which might be due to synchrotron X-rays from a separate
population of electrons or of protons \citep{Aharonian02}.
Detection
of two-sided X-ray jets in the FR II radio galaxies Cyg~A
\citep{Wilson00,Wilson01}, 3C353 \citep{Kataoka08}, and Pictor A
\citep{Hardcastle16} indicate a Doppler factor around unity that does
not allow an iC/CMB origin.  The optical polarization in the jet of
the quasar PKS 1136-135 indicates production via synchrotron emission,
and is best explained as arising from the low-energy tail of the X-ray
emitting population \citep{Cara13}.  \citet{Meyer15,Meyer17} use upper
limits to \emph{Fermi} $\gamma$-ray emission and also ALMA imaging of
the jet in the quasar PKS 0637-752 to construct models that do not
allow iC/CMB emission to explain the X-rays, providing that the ALMA
and optical emission are the high energy extension of the radio
synchrotron spectrum. The complex structure of the jet in the quasar
PKS1127-145 requires at least two emission components, which may
include both iC/CMB and synchrotron components
\citep{Siemiginowska07}.

Despite these challenges, at sufficiently large redshifts the CMB
energy density must dominate over  magnetic energy density,  and
iC/CMB X-rays will 
result \citep{Schwartz02}. High redshift X-ray jets have been reported
\citep{Siemiginowska03b,Cheung06,Cheung12,McKeough16}, most remarkably the jet
in the {\it z}=2.5 quasar B3 0727+409 for which the only radio
detection is single knot 1\farcs4 from the core. No further extended  radio
emission was detected along the $\approx$ 10\arcsec\ long X-ray 
jet \citep{Simionescu16}.  \citet{Lucchini17} suggest
that cooling of the highest energy electrons can result in X-ray jets
that are "silent in the radio and optical bands."
These considerations motivate continued efforts to test the
iC/CMB model at lower redshifts.

The primary purpose of this paper is to present the broadband data
collected on the 4C+19.44 jet ({\it z}=0.72). We will show that a consistent
interpretation in terms of the iC/CMB mechanism is possible for the
bulk of the X-ray jet. If the iC/CMB scenario is ultimately proven, it
provides a means to deduce the otherwise unobservable low energy tail
of the electrons producing GHz radiation. That low energy tail
contains the bulk of the relativistic energy budget of the emitting particles, and must be
estimated in order to apply minimum energy or equipartition arguments
to measure the magnetic field.

4C+19.44 (=PKS 1354+195) was included in a \chan\ and HST survey
project \citep{Sambruna02,Sambruna04,Marshall05} that was based on a
selection of radio jets that were asessed as having high probability
of detection by \chan\ in a 5 to 10 ks 
observation.  We selected this source for longer observations because
the 10 ks \chan\ observation demonstrated that the entire jet to the
South of the quasar was detected in the X-rays and because two inner
knots were also optically-detected with the Hubble Space Telescope
(HST) \citep{Sambruna02}. Preliminary results from these longer \axaf\
observations have been reported \citep{Schwartz07,Schwartz07b}.  In
addition to the deep \chan\ observations, we obtained HST observations
(475 and 814 nm), a Spitzer image (3.6 $\mu$m) and data at three radio
frequencies (1.4, 5, and 15 GHz) with the NRAO\footnote{The National
  Radio Astronomy Observatory is a facility of the National Science
  Foundation operated under cooperative agreement by Associated
  Universities, Inc.} Very Large Array (VLA).  With many resolution
elements down the jet, our primary goal was to evaluate the spectral
energy distributions as a function of distance from the quasar in
order to constrain the emission processes for the various bands.

We adopt $h={\rm H}_{0}$/(100 km s$^{-1}$ Mpc$^{-1}$)=0.67, $\Omega_{\rm
M}=0.27$ and $\Omega_{\rm \Lambda}=0.73$, so that at a redshift of
0.719 \citep{stei91} 1\asec\ corresponds to 7.7 kpc.  Spectral
indices, $\alpha$, are defined by flux density S$_{\nu} \propto
\nu^{-\alpha}$.

\section{The Data\label{sec:obs}}

\subsection{\chan\ X-ray Data}

Our deep \chan\ observation was scheduled as four separate pointings
in 2006 \\
\dataset[\chan\ ObsIDs
 6903, 6904,  7302, and 7303]{http://cda.harvard.edu/chaser?obsid=6903,6904,7302,7303}
for a total of 199 ks on target as summarized in
Table~\ref{tab:xobs}.  We observed using only the back illuminated
ACIS chip S3 in a 1/4 sub-array mode to reduce the effects of pile-up of
the bright nucleus. This results in a dead-time fraction of about 5\%
(0.04104 s readout time divided by the 0.84104 frame time) for a net
observation live-time of 189.35 ks.  All data were obtained with ACIS-S
in the faint mode; i.e., telemetering the 3$\times$3 pixel amplitudes.
A range in roll angle was requested to position the CCD
charge-transfer readout streak away from the jet. 
ObsID 6904 gave about 35 ks live-time taken at roll angle 120\degr ,
while the remaining observations were all at 137\degr .  Although we
encountered a star tracker problem during the first half of ObsID 7302
that produced a displacement to the east, the offset was of order 0\farcs2
 so for the purposes of photometry, we did not reject these
data. Previous results were reported using CALDB 3.2.1; more recently
we have reanalyzed all the data 
using CALDB 4.5.1.1 and CIAO 4.6.  These give an appropriate ACIS
contamination model, and use the energy dependent sub-pixel event
redistribution algorithm (EDSER).  Various members of our team have
analyzed the data independently. We also add the reprocessed data from
\dataset[\chan\ ObsID 2140]{http://cda.harvard.edu/chaser?obsid=2140}
originally published by
\citet{Sambruna02,Sambruna04}, to give a total live-time of 198.4 ks.

\begin{table}
\begin{center}
\caption{Summary of the  \chan\ X-ray Observations\label{tab:xobs}}
\begin{tabular}{crrr}
\tableline\tableline
Observation Date & ObsID & Live Time & Roll Angle\\
 & & (ks)\\
\tableline
2001 Jan 08 & 2140\tablenotemark{a} & 9.056 &  66\degr\\
2006 Mar 20 & 6904\tablenotemark{b}& 34.958 & 120\degr\\
2006 Mar 28 & 7302\tablenotemark{b}& 68.936 & 137\degr\\
2006 Mar 30 & 7303\tablenotemark{b}& 41.523 & 137\degr\\
2006 Apr 01 & 6903\tablenotemark{b}& 43.933 & 137\degr\\
\tableline
\end{tabular}
\end{center}
\tablenotetext{a}{\citet{Sambruna02,Sambruna04}}
\tablenotetext{b}{This paper}
\end{table}

An alternate analysis was reported in \citet{Massaro11}. They created  flux maps  in the soft
(0.5--1 keV), medium (1--2 keV), and hard (2--7 keV) bands. For each
band, the data were divided by the exposure map and multiplied by the
nominal energy of each band, resulting in maps with units of erg~cm$^{-2}$~s$^{-1}$.
Photometry was then performed for each region using
funtools\footnote{https://github.com/ericmandel/funtools} and is
reported in the on-line version of Table 7 of \citet{Massaro11} for
ObsID 7302.

\begin{deluxetable}{llll}
\tablewidth{0pc}
\tablecaption{Photometric Regions\label{tab:regions}}
\tablehead{
\colhead{Description} & \colhead{shape}   & \colhead{Position\tablenotemark{a}}    &
\colhead{Size\tablenotemark{b}}\\
 &  &  \colhead{(J2000.0)} &   
}
\startdata
N16.0; N hot spot  & box & 13:57:04.174,+19:19:22.54 & 2\farcs46,1\farcs72,163$^{\circ}$\\
N15.4; N lobe  & circle & 13:57:03.959,+19:19:21.06 & 6\farcs4\\
S2.1   & box & 13:57:04.476,+19:19:05.36 & 2\farcs12,1\farcs72,163$^{\circ}$\\
S4.0   & box & 13:57:04.507,+19:19:03.57 & 1\farcs54,1\farcs72,163$^{\circ}$\\
S5.3   & box & 13:57:04.531,+19:19:02.30 & 1\farcs08,1\farcs72,163$^{\circ}$\\
S6.6   & box & 13:57:04.540,+19:19:00.97 & 1\farcs52,1\farcs47,163$^{\circ}$\\
S8.3   & box & 13:57:04.583,+19:18:59.37 & 1\farcs87,1\farcs72,163$^{\circ}$\\
S10.0  & box & 13:57:04.613,+19:18:57.81 & 1\farcs38,1\farcs72,163$^{\circ}$\\
S11.2  & box & 13:57:04.661,+19:18:56.65 & 1\farcs23,1\farcs72,163$^{\circ}$\\
S12.9  & box & 13:57:04.697,+19:18:55.20 & 1\farcs87,1\farcs72,163$^{\circ}$\\
S14.6  & box & 13:57:04.732,+19:18:53.58 & 1\farcs47,1\farcs72,163$^{\circ}$\\
S15.9  & box & 13:57:04.760,+19:18:52.29 & 1\farcs23,1\farcs72,163$^{\circ}$\\
S17.7  & box & 13:57:04.800,+19:18:50.50 & 2\farcs46,1\farcs72,163$^{\circ}$\\
S21.6; transition region  & box & 13:57:04.799,+19:18:46.37 &
5\farcs4,5\farcs4,163$^{\circ}$\\ 
S25.7; entrance SHS  & circle & 13:57:04.887,+19:18:42.52 & 1\farcs33\\
S28.0; S hot spot  & circle & 13:57:04.780,+19:18:39.89 & 1\farcs72\\
North Background&box&13:57:03.568,+19:19:55.93&35\farcs04,49\farcs35,0$^{\circ}$\\
Southwest Background&box&13:57:03.422,+19:18:24.28,&24\farcs26,24\farcs05,315$^{\circ}$\\
\enddata

\tablenotetext{a} {Center of the box or circle.}
\tablenotetext{b}{Size of boxes given as length, width, position angle
    counter clock-wise from North. For circles, size is the radius.}
\end{deluxetable}

\subsubsection{Photometric Regions \label{sec:photo}}

\begin{figure} 
\begin{center}
\includegraphics[
height=5.in]{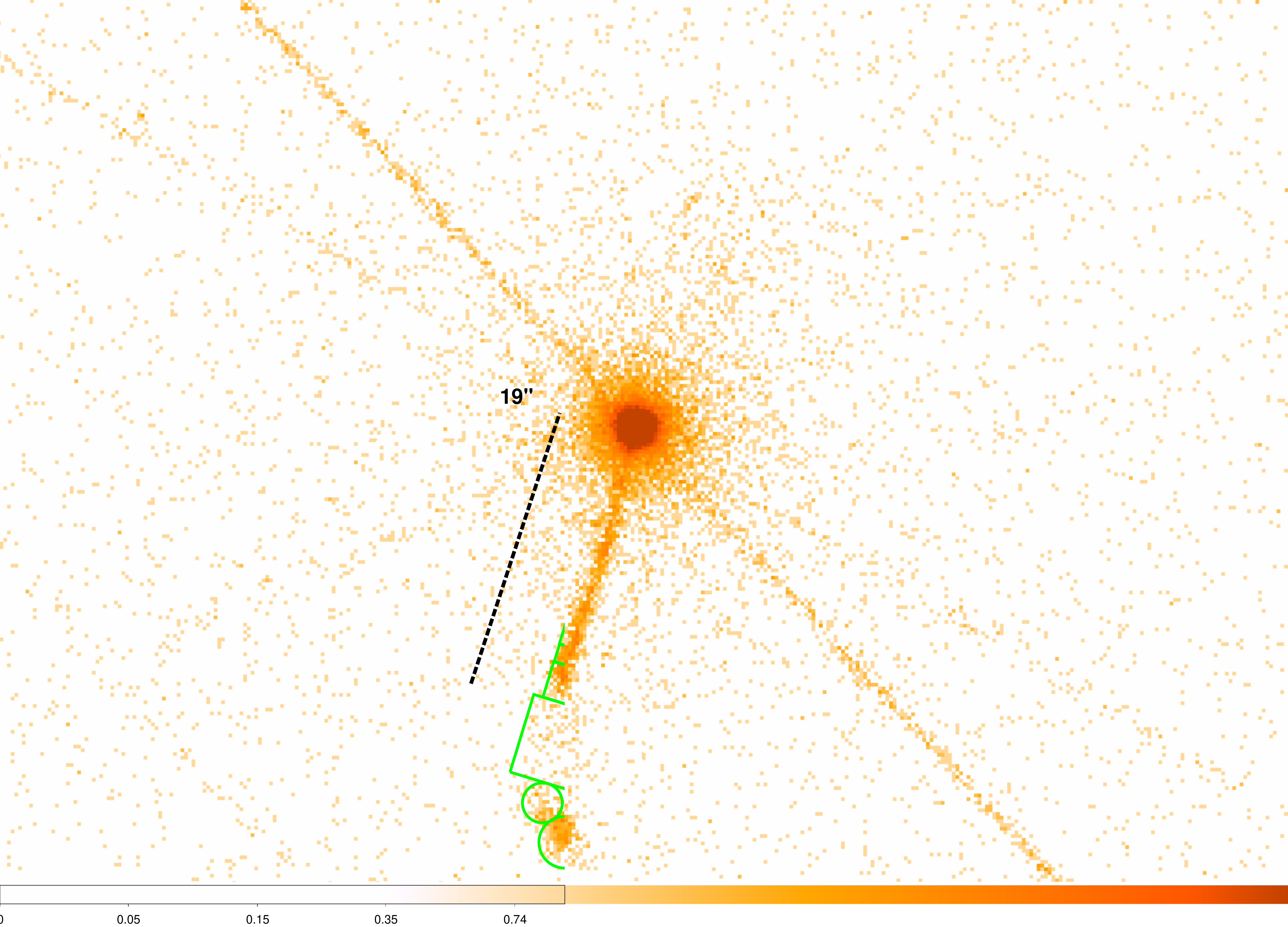}
\end{center}
\caption{The regions defined by the X-ray morphology, and used for photometry
  are delineated by the solid circles and rectangles defined in
  Table~\ref{tab:regions}.
   The figure  is a summed X-ray count map (0.5--7 keV) binned in
   0\farcs246 $\times$ 0\farcs246 pixels. From top (North) to bottom,
   N16.0 is the small 
   rectangle for the northern hot-spot, and N15.4 is  the large circle for the
   northern   lobe. The bright and the faint streaks from NE to SW are
   the CCD readout
  streaks from the observations at the two different roll angles. 
  The  jet is  composed of the 11 rectangular regions starting
  2\farcs1 S of the quasar 
  nucleus and  ending at S17.7.  The large rectangle, S21.6 is the
  `transition region' which has   only low surface brightness in both
  the radio and X-ray bands. The  next region is a
  small circle, S25.7 which is the `entrance' to the S   hot-spot, and
  the last, larger circle is the S hot-spot. Numbers in the name of
  each region refer to the center of the box or circle. The color bar
  gives counts per pixel, and saturates at 50 counts per 0.0605
  arc-second$^2$. 
  \label{fig:regs}}
\end{figure}

Since our primary interest is to determine the spectral energy
distribution (SED) for each X-ray feature, we defined our regions on
the basis of the X-ray morphology.  The regions are shown in
figure~\ref{fig:regs} and are labeled with their direction (N or S)
and distance in arc-seconds of the region center from the quasar,
following the convention defined in \citet{Schwartz00}.  Each ObsID
was adjusted by an amount between 0\farcs21 and 0\farcs30 to superpose
the quasar core at the position given in \citet{Johnston95};
RA:13$^{\rm h}$57$^{\rm m}$04.4366$^{\rm s}$ and
DEC:+19$^\circ$19\arcmin 07\farcs372
~(J2000). N16.0 is a rectangle for the northern hot-spot and N15.4 is a
large circle to deal with the northern lobe.  Moving south from the
quasar, the first region is designated S2.1 and there are a total of
11 rectangles along the jet.  S6.6 is not centered on the jet: the
eastern 0\farcs25 has been trimmed off in order to avoid optical
emission associated with a (foreground?) edge-on galaxy.  After the
main jet, there is a large square, S21.6 which is termed `the
transition region'.  It contains low brightness emission in both X-ray
and radio bands: although the morphology is not well defined, the
emission may arise from the southern lobe.  Finally there are two
circular regions, S25.7, called `the entrance to the hot-spot' and
S28.0, the southern hot-spot itself.  The regions are shown in
Figure~\ref{fig:regs} and specified  in
Table~\ref{tab:regions}.  Detector background was determined from two
large rectangular regions, north of the northern lobe and southwest of
the southern lobe. These are not shown, but are specified in
Table~\ref{tab:regions}. Table~\ref{tab:fluxes} gives the resulting 1
keV flux densities for all the regions, along with those at other
frequencies as discussed in sections~\ref{hst},~\ref{irac},
and~\ref{vla}.

\subsubsection{Spectral Analysis}\label{sec:xspec}

Modeling of the X-ray emission from several regions has been performed
in Sherpa \citep{freeman2001} version CIAO 4.6. We extracted the
spectra and created response files for each observation and used the
energy range 0.5--7~keV for all the spectral modeling.  The number of
net counts from the jet in each region, used for the
photometry reported in Table~{tab:fluxes}, ranged from 39.3
to 184.5, after subtracting the detector background and scattered
photons from the quasar itself. We determined the latter from a Marx
5.1 simulation of the quasar, using the quasar's measured spectral
energy index, $\alpha_{\rm x}$, of 0.66 \citep{Marshall17}, and
incorporating pile-up, the ACIS readout streak, and EDSER.  In region
S2.1 the quasar can account for the entire signal, leaving the ten
regions from S4.0 to S17.7 for analysis of the jet.  For spectral
fitting we neglected the 
background counts, predicted to range from 1.2 to 2.6, i.e., less than
2\% of the total counts in any region. Scattered photons from the
quasar give 44\% and 39\% of the counts in regions S4.0 and S5.3, but
should bias the fit to the index by less than the estimated errors,
since the quasar has a similar spectrum. We used the
Nelder-Mead optimization algorithm and Cash likelihood statistics
appropriate for low counts and fit the data in Sherpa.

For the spectral analysis we use only the four contemporaneous ObsID's
from 2006 (Table~\ref{tab:xobs}). We fit the same model jointly to the four
individual spectra. 
We assumed an absorbed power law model for each region, with the
absorption column frozen at the Galactic value of $\rm N_{H} = 2.23\times
10^{20} $~cm$^{-2}$. No absorption in excess of Galactic was detected,
with 90\% upper limits for an absorber at redshift ${\it z}=0.5$ with a
range of $(0.2$ to $0.4) \times 10^{22} $~cm$^{-2}$ for the different
regions.  
We then froze the excess absorption at zero, and fit
the photon index $\Gamma$ of a power law and the
normalization. We get the same results using {\it XSPEC} or {\it
  Sherpa}. Table \ref{tab:xspec} lists the X-ray spectral
results, reporting the energy index $\alpha_{\rm x} = \Gamma -1$, along
with the radio spectral index as discussed in section~\ref{radiojet}.

\begin{figure}
\includegraphics[
height=7.5in]{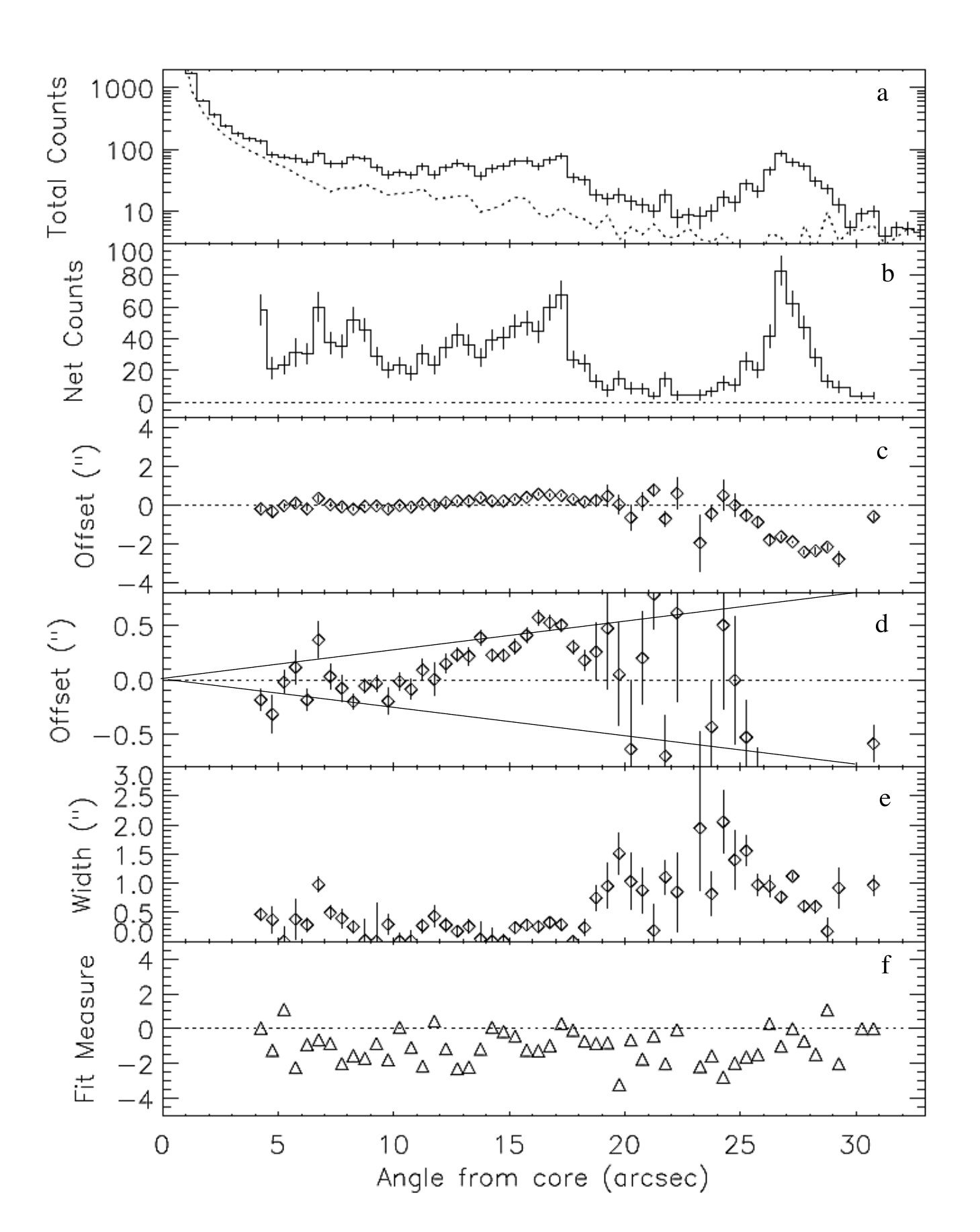}
\caption{Panel \textit{a}: solid histogram gives \axaf\ X-ray counts
  per bin along the jet through the southern hotspots. Dashed
  histogram gives counts per bin in a region 180\degr\ from the jet.
  Panel \textit{b}: net counts, subtracting the region
  180\degr\ away. Panel \textit{c}: the deviation of the centroid of
  each bin from the mean position angle of 165\degr . Panel \textit{d}
  is the same data plotted at a stretched scale to emphasize the inner
  straight 17\arcsec\ jet, and with the solid lines indicating a
  deviation of $\pm$1\fdeg5 from 165\degr . Panel \textit{e}: The
  width, defined as the difference in quadrature of a fit to the
  observed data plus background, and the intrinsic resolution
  $\sigma=0\farcs34$ determined from the readout streak. Panel
  \textit{f}: 
  the fit measure is calculated from the probability determined from a
  Kolmogorov-Smirnov goodness of   
  fit test, converted to a Gaussian deviate giving the same probability.
\label{fig:hlmprofiles}}
\end{figure}

\subsubsection{X-ray Jet Structure}

We generated profiles of the X-ray jet, using likelihood-based
Gaussian fitting with Poisson statistics. We sorted the summed counts
from 0.5 to 7 keV into 0\farcs5
bins along the jet, taking  counts $\pm$5\arcsec\ perpendicular
to the jet. 
Figure~\ref{fig:hlmprofiles} plots the data  along the jet as a function
of distance from the quasar core. Each fit was to a one dimensional
Gaussian normal to the jet, with free parameters being the total
number of counts, the position relative to the mean position angle of
165\degr, the sigma of the Gaussian, and the background level data
from a strip at position angle 345\degr. Small aspect residuals were
reduced by fitting the quasar core in right ascension and declination
in separate 300 s time intervals, and fitting the residuals from the
known position to a polynomial to correct the data. Fitting a Gaussian
shape across the ACIS
readout streak gives $\sigma$= 0\farcs34, and represents the
intrinsic response to an unresolved line source. Subtracting this number in
quadrature from the standard deviation gives a measure of the intrinsic width
of the jet, plotted in the second panel up from the bottom. 

Closer than 3\farcs5 from the quasar the jet parameters cannot be
determined.  The width is marginally resolved in the
3\farcs5 to 17\arcsec\ region, but contaminated due to the galaxy
SDSS~J135704.63+191900.9 which has a significant X-ray flux density
$\approx$0.08 nJy, and overlaps the jet  in the 6\arcsec\ to 7\arcsec\
region.  The X-ray emission is clearly 
detected and very broad between 20\arcsec\ and 25\arcsec. This region
bridges the straight part of the jet and the terminal hot-spot, and is
probably part of a lobe structure. Within the 17\arcsec\ jet there are
some significant, but small, offsets of the position angle of the jet
center line (middle panels
of Figure~\ref{fig:hlmprofiles}).

\subsection{HST Data\label{hst}} 
 
\begin{figure*}[t]
\includegraphics[width=0.98\columnwidth]{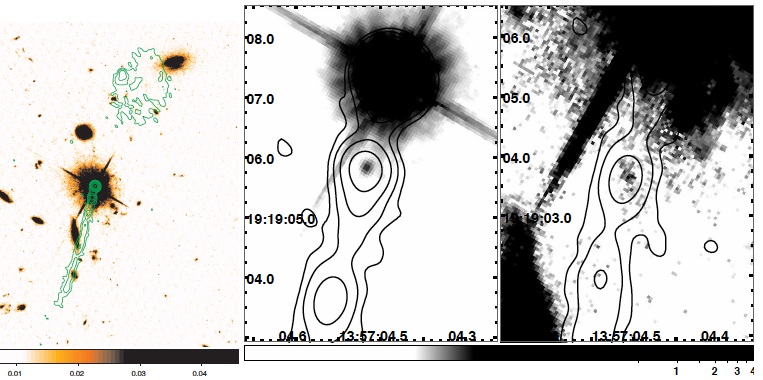}
\caption{Left panel: The HST F475W image with 1.4 GHz radio contours
  overlaid. Middle panel: HST F814W image with 5 GHz contours from the
  the VLA A array overlaid. The grey scale mapping is adjusted to show
  the knot S2.1. Right panel: Same data as the middle panel, but with
  the field of view and grey scale adjusted to show the detection of
  S4.0. S5.3 is also detected in all three HST filters. The lowest
  radio contour is 0.25 mJy/beam with contours increasing by factors
  of 4.\label{fig:opt3pan}}
\end{figure*}

Our HST data were obtained with the WFC-ACS (proposal ID 10762) on
2006 March 23 (F814W) and March 24 (F475W).  Exposure times were 6998 s
and 4472 s respectively. We also included analysis of some archival
WFPC2 F702W data from 1996 Jun 22 (4600 s exposure; proposal ID 5984).
The images were processed in the usual manner with CAL\_VER 4.6.1.
The 8060\AA\ image is shown in Figure~\ref{fig:opt3pan}.

The left panel shows the field around the quasar. The presence
of many galaxies of similar size and magnitude (particularly to the
East of the quasar) is suggestive of a group or cluster.
\citet{Ellingson91} searched for a cluster associated with 4C+19.44
and studied seven galaxies within 2\arcmin\ of which four had measured
redshifts between 0.36 and 0.53.  The NASA/IPAC Extragalactic Database
(NED) indicates that 4C+19.44 has absorption line systems at {\it z} =
0.431, 0.457, and 0.522 \citep{ryab03}. There are three galaxies
listed by NED that lie within 30$^{\prime\prime}$; these have
spectroscopic redshifts in the range {\it z} = 0.43-~0.46, while the
SDSS-measured redshift of 4C+19.44 is 0.7196 \citep{Schneider10}.
There are no other objects within 4\arcmin\ that have SDSS spectra.
Attempts to confirm either a foreground cluster or a cluster
associated with the quasar using SDSS photometric redshifts were
inconclusive. The measured X-ray profile in an azimuthal sector of
70\degr\ to the West tracks the profile of the simulated point quasar
plus background, and puts a 2$\sigma$ upper limit of
2$\times$10$^{43}$ erg cm$^{-2}$ s$^{-1}$ for emission from an assumed
cluster with temperature 2 keV at the redshift of the quasar.

The other two panels of Figure~\ref{fig:opt3pan} have had their
contrasts adjusted to emphasize the optical detections of knots within
regions S2.1 and S4.0. We find that S2.1 has an
apparent diameter of 0\farcs2, consistent with the deconvolved radio
major axis, whereas S4.0 has an extent along the jet of
$\approx$ 0\farcs4, again consistent with the radio size in the PA of
the jet.  Emission from the region S5.3 is also significant although
barely visible in Figure~\ref{fig:opt3pan}. The HST photometry
reported in Table~\ref{tab:fluxes} was performed on images that had a
first order subtraction of the quasar.

\begin{deluxetable}{rrrrrrrl}
\tabletypesize{\scriptsize}
\tablewidth{0pc}
\tablecaption{Flux densities for X-ray defined regions\label{tab:fluxes}}
\tablehead{
\colhead{Region} & \colhead{4.86 GHz}   & \colhead{14.9 GHz}    &
\colhead{3.6 $\mu$m}    & \colhead{F814W}    & \colhead{F702W}  & \colhead{F475W}    &
 \colhead{1.0 keV}    \\
\colhead{}  &  \colhead{4.9~10$^9$ Hz} & \colhead{1.5~10$^{10}$ Hz} & \colhead{8.3~10$^{13}$ Hz} & \colhead{3.7~10$^{14}$ Hz} & \colhead{4.3~10$^{14}$ Hz} &
    \colhead{6.3~10$^{14}$ Hz} & \colhead{2.4~10$^{17}$ Hz} \\
\colhead{}      &  \colhead{(mJy)}    &  \colhead{(mJy)}    &
\colhead{($\mu$Jy)}    &  \colhead{($\mu$Jy)}   &  \colhead{($\mu$Jy)}  &  \colhead{($\mu$Jy)}    & 
\colhead{(nJy)}   
}
\startdata
N16.0  & 61.1$\pm$0.30 & 23.7$\pm$0.4 & \nodata &$<$0.52 &$<$0.24  & $<$0.12  & 0.101$\pm$.028 \\
N15.4  & 173$\pm$20 & 48.2$\pm$2.5 & \nodata & $<$15 &$<$6.2  &$<$3.8  & 0.822$\pm$0.111 \\
S2.1    & \nodata & \nodata & \nodata & 0.57$\pm$0.09\tablenotemark{a} & 0.40$\pm$0.15\tablenotemark{a} & 0.21$\pm$0.07\tablenotemark{a}  & $<$0.42\tablenotemark{b} \\
S4.0   & 16.5$\pm$0.20 & 7.21$\pm$0.36 & \nodata &
0.11$\pm$0.01\tablenotemark{a} &   0.07$\pm$0.04\tablenotemark{a}  &
0.07$\pm$0.02\tablenotemark{a}   & 0.513$\pm$0.079 \\ 
S5.3   & 6.69$\pm$0.21 & 2.92$\pm$0.28 & \nodata &
0.07$\pm$0.03\tablenotemark{a}&  0.08$\pm$0.03\tablenotemark{a}   &
0.04$\pm$0.01\tablenotemark{a}   & 0.184$\pm$0.045\\ 
S6.6   & 11.0$\pm$0.20 & 4.57$\pm$0.33 & $<$10 & $<$0.28  & $<$0.26&
$<$0.11 & 0.396$\pm$0.051  \\ 
S8.3   & 19.0$\pm$0.30 & 7.56$\pm$0.39 & $<$10 & $<$2.59 & $<$1.64 &
$<$0.33  & 0.693$\pm$0.062 \\ 
S10.0  & 6.19$\pm$0.23 & 2.29$\pm$0.34 & $<$10 & $<$1.74 & $<$0.59  &
$<$0.84 & 0.302$\pm$0.042 \\ 
S11.2  & 5.34$\pm$0.22 & 2.00$\pm$0.32 & $<$6 & $<$0.14 & $<$0.07&
$<$0.10 & 0.310$\pm$0.041  \\ 
S12.9  & 9.91$\pm$0.27 & 4.11$\pm$0.39 & $<$12 & $<$1.94 & $<$1.05 &
$<$0.55 & 0.643$\pm$0.058 \\ 
S14.6  &9.18$\pm$0.24  & 3.48$\pm$0.35 & $<$6 & $<$0.10 & $<$0.23 &
$<$0.10 & 0.604$\pm$0.054 \\ 
S15.9  & 2.79$\pm$0.22 & 1.10$\pm$0.32 & $<$6 & $<$0.10& $<$0.12 &
$<$0.08 & 0.581$\pm$0.053\\ 
S17.7  &2.48$\pm$0.31   & 0.87$\pm$0.45 & $<$6 & $<$0.12 & $<$0.10&
$<$0.06  & 0.865$\pm$0.065\\ 
S21.6\tablenotemark{c}  & 5.45$\pm$0.82 & 3.01$\pm$1.18  & \nodata &
$<$1.1 & $<$0.83 & $<$0.52 & 0.434$\pm$0.056 \\ 
S25.7\tablenotemark{d}  & 3.17$\pm$0.35 & 1.50$\pm$0.51 & $<$6 &
$<$0.33  & $<$0.46 & $<$0.08 & 0.392$\pm$0.045 \\ 
S28.0\tablenotemark{d} & 85.7$\pm$0.5 & 30.4$\pm$0.7 & $<$6 & $<$0.60
& $<$0.52 & $<$0.38 & 1.09$\pm$0.074  \\ 
\enddata

\tablecomments{X-ray flux densities were derived from the observed
  fluxes assuming 
an energy index $\alpha_{\rm x}$=0.8.}  

\tablenotetext{a}{Knots  S2.1, S4.0 and S5.3 have actual HST detections,
  with a resolved size $\approx$ 0.2$^{\prime\prime}$. Other regions
  list a 2$\sigma$ upper limit. Bright galaxies to the south cause larger upper limits.}

\tablenotetext{b}{The X-ray photometry of S2.1 is compromised by the
PSF of the quasar. A 2$\sigma$ upper limit is quoted.}%

\tablenotetext{c}{S21.6 is a large area of low brightness.  It may
  well be a lobe; it is not a normal part of the jet.}

\tablenotetext{d}{S25.7 is a bit of emission entering the south hot-spot
  and S28.0 is the south hot-spot.}

\end{deluxetable}

\subsection{Spitzer IRAC Data\label{irac}}

Our \emph{Spitzer} IRAC data were taken on 2005 July 16 as part of our
Cycle-1 General Observer program  \citep{Uchiyama06}.  We
chose the pair of 3.6 and $5.8\ \mu\rm m$ arrays for best spatial
resolution and sensitivity.  The native pixel size in both arrays is
$\simeq 1\farcs22$, and the point spread functions are $1\farcs66$ and
$1\farcs88$ (FWHM) for the 3.6 and $5.8\ \mu\rm m$ bands,
respectively.  We obtained a total of 60 frames per channel, each with
a 30-s frame time, and the frame data were combined into a single
image using the Spitzer Science Center  software {\it mopex}.  We
subtracted the PSF wing 
of the bright quasar core from each IRAC image.  Also, some field
galaxies (see the \emph{Hubble} image) near the southern jet were
subtracted.

\begin{figure}
\includegraphics[width=6.in]{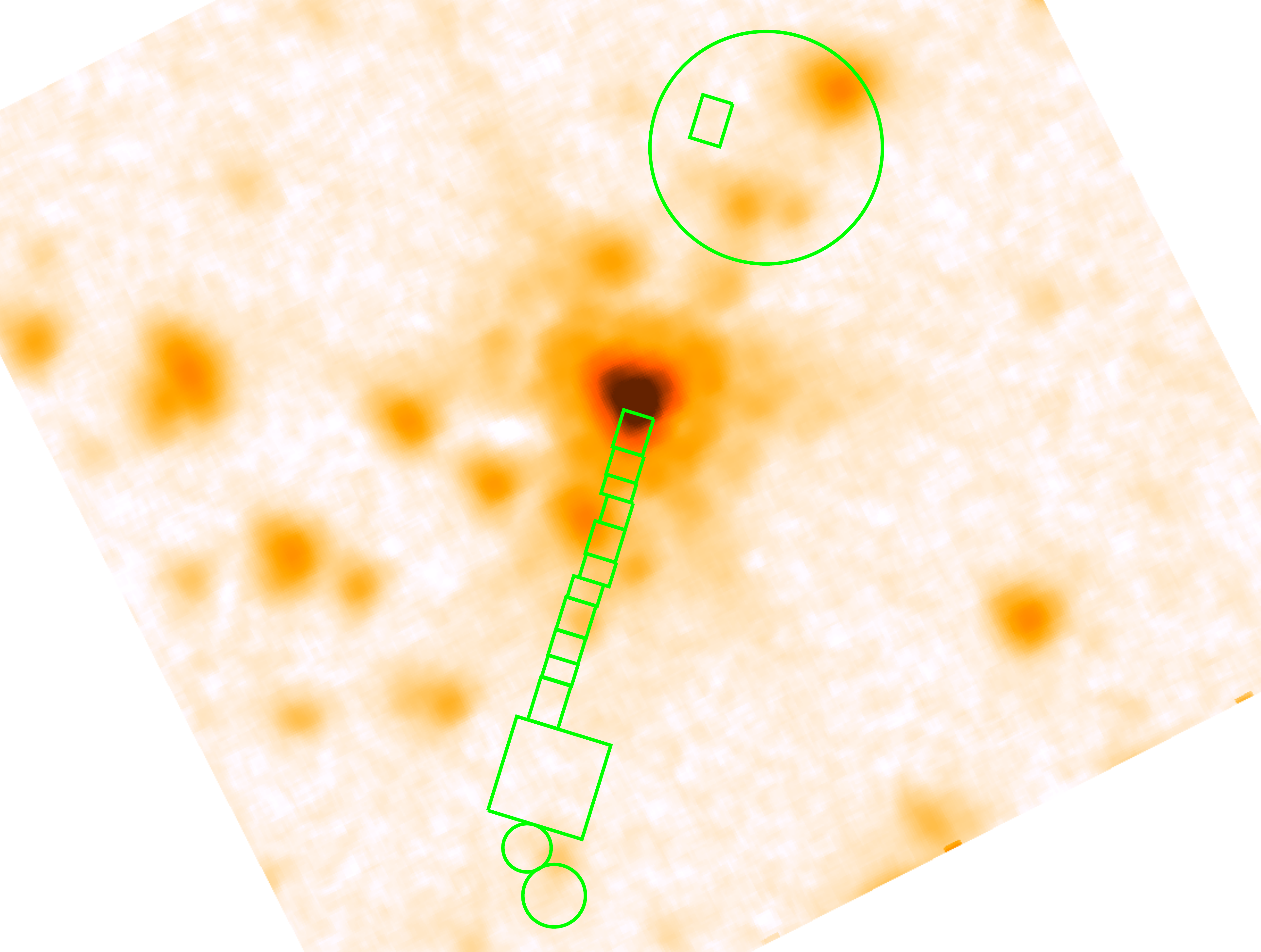}
\caption{Spitzer IRAC 3.6 micron image, with the X-ray photometry
  regions overlaid. All apparent emission can be attributed to the
  quasar and foreground galaxies. The jet surface brightness is too
  low to detect with the angular resolution of Spitzer. \label{fig:irac}}
\end{figure}

No significant infrared emission was found along the jet with the
\emph{Spitzer} IRAC (Figure~\ref{fig:irac}).  Based on statistical fluctuations of the surface
brightness of the core-subtracted IRAC image and the uncertainties
associated with PSF removal (adopting 10\% of the quasar's PSF wing
intensity at the location being considered), we place a $3\sigma$
upper limit on the $3.6\ \mu\rm m$ flux of a point source in this
region as $f_{3.6} = 6\ \mu\rm Jy$.  As reported in 
Table~\ref{tab:fluxes}, we adopt this value as the upper limit for
each jet knot and hot-spot except that we place more conservative
limits for some regions that contain field galaxies (S6.6, S8.3,
S10.0, and S12.9).  We do not report the upper limits for the $5.8\
\mu\rm m$ band because they are much less constraining.

\subsection{VLA Observations\label{vla}}

\begin{figure}
\epsscale{0.8}
\plotone{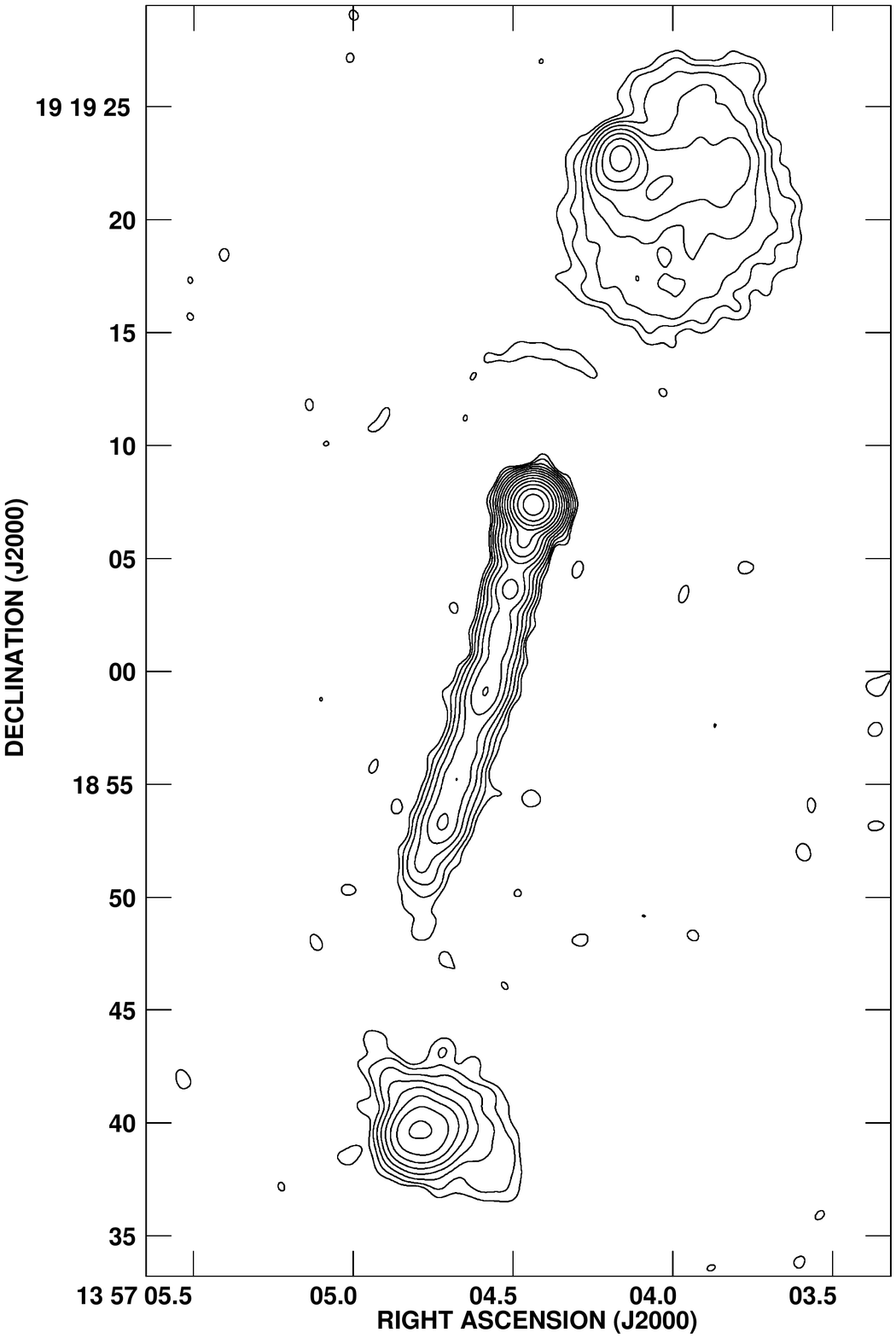}
\caption{VLA image at 4.86 GHz (B-array).  The parameters of the image are
given in Table \ref{tab:radobs}; contours increase by factors of 2 and
the lowest contour level is 0.2 mJy/beam. \label{fig:radmap}}
\end{figure}

We have performed radio observations of the quasar with the
NRAO VLA (program S71062) at two
epochs: 2006 February 06 at 1.4 and 4.86~GHz with A-array and 2006
July 30 at 4.86 and 14.965~GHz with B-array.  The data were reduced in
the usual manner with the Astronomical Image Processing System (AIPS)
using 1407+284 as a phase and polarization D-terms calibrator
and 3C~286 as an amplitude and polarization position angle
calibrator. We also use {\it Difmap} for modeling calibrated uv-data.
We have obtained images for three Stokes parameters, I, Q, and U. At
1.4~GHz we have imaged separately the two intermediate frequency
bands centered on 
1.365~GHz and 1.435~GHz. Table~\ref{tab:radobs} summarizes the radio
observations.  The highest dynamic-range image of the quasar is
presented in Figure~\ref{fig:radmap}, which shows that the morphology
of the radio source is similar to the X-ray structure: a bright
compact core, a prominent jet to the south-east up to 17\asec ~from
the core at position angle $\sim$165$^\circ$, a southern hot spot with
faint diffuse emission, and a diffuse northern lobe with a
hot spot. We have measured the flux at all wavelengths in the
photometric regions described in section~\ref{sec:photo}  using the AIPS task
IMEAN. The measurements at 4.86 GHz and 14.9 GHz are listed in
Table~\ref{tab:fluxes}, for the regions defined in
Table~\ref{tab:regions}. At 1.4 GHz the beam size is large enough that
some of the flux density ``spills'' out of the smaller photometric
regions.

\begin{figure*}[t]
\epsscale{0.9}
\plotone{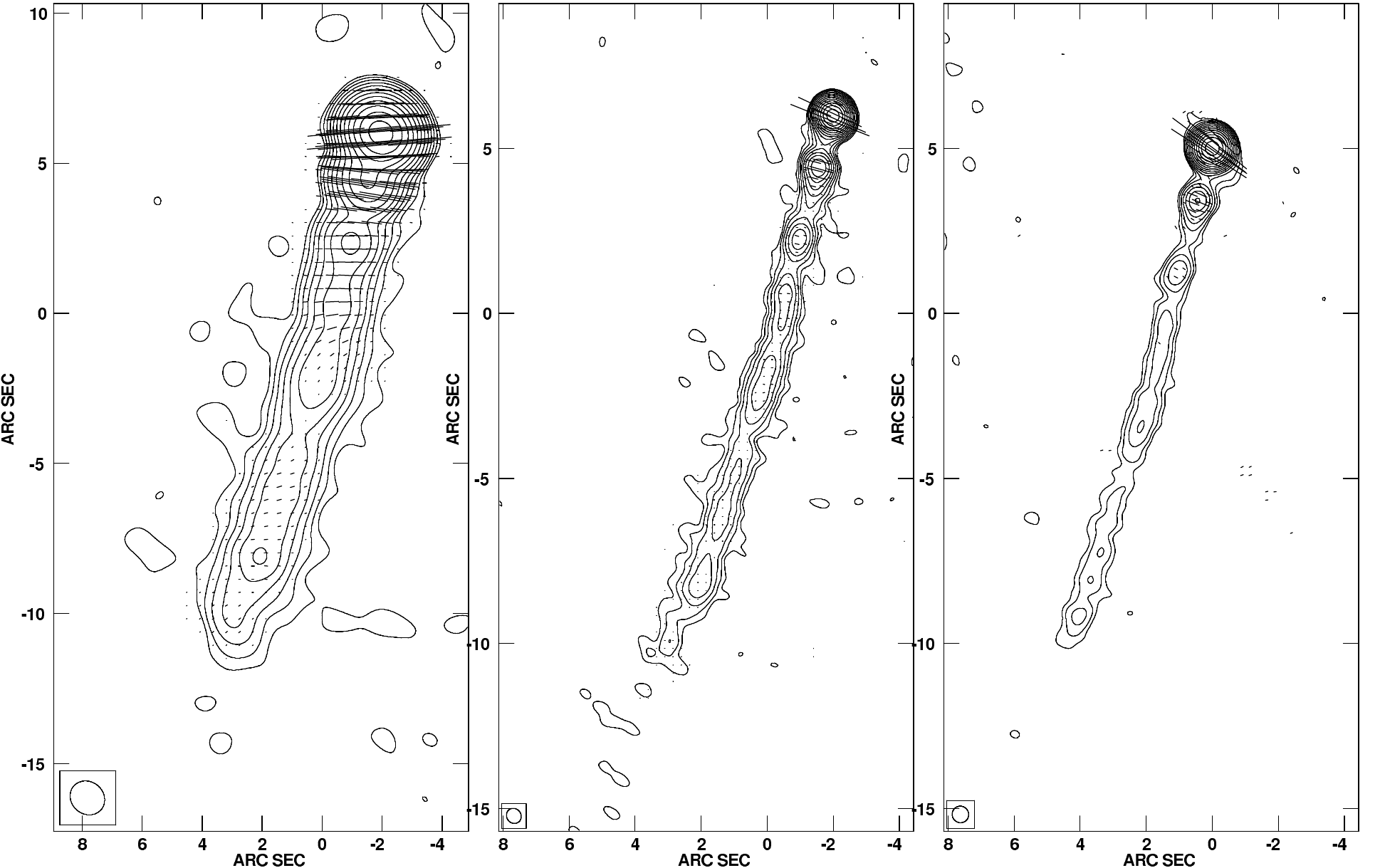}
\caption{VLA images at different wavelengths: {\it left} - at
  1.365~GHz (A-Array, lowest contour is 0.45 mJy/beam), {\it middle} - at
  4.86~GHz (A-Array, lowest contour is 0.2 mJy/beam), and {\it right}
  - at 14.965 ~GHz (B-Array, lowest contour is 0.3 mJy/beam). The parameters of
  the image are given in Table \ref{tab:radobs}; contours increase by
  factors of 2; linear segments show polarization E-vectors; the
  length of vectors is proportional to the local polarized
  intensity. \label{fig:jetmaps}}
\end{figure*}

\begin{figure}
\includegraphics[width=6.in]{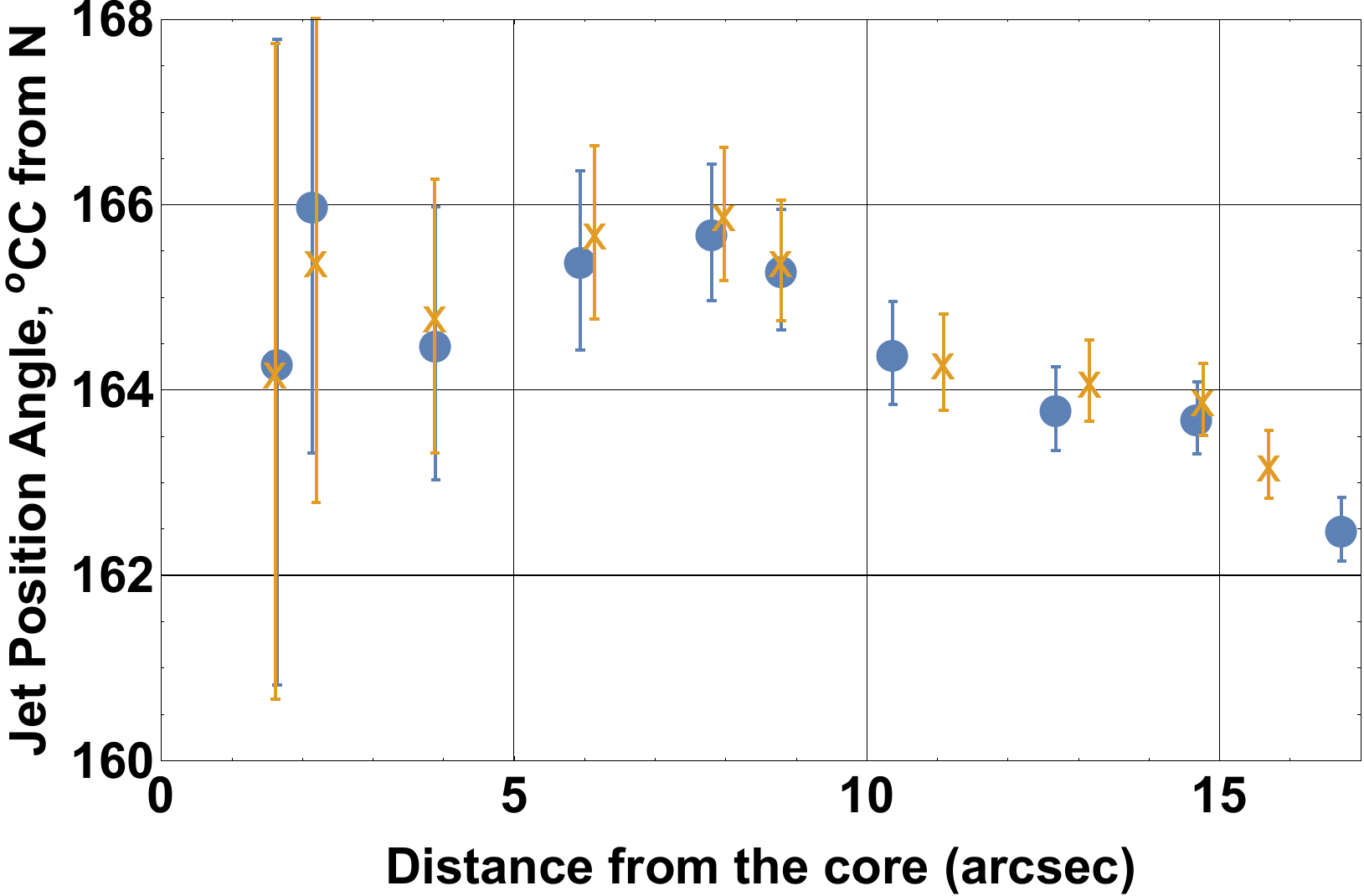}\\
\includegraphics[width=6.in]{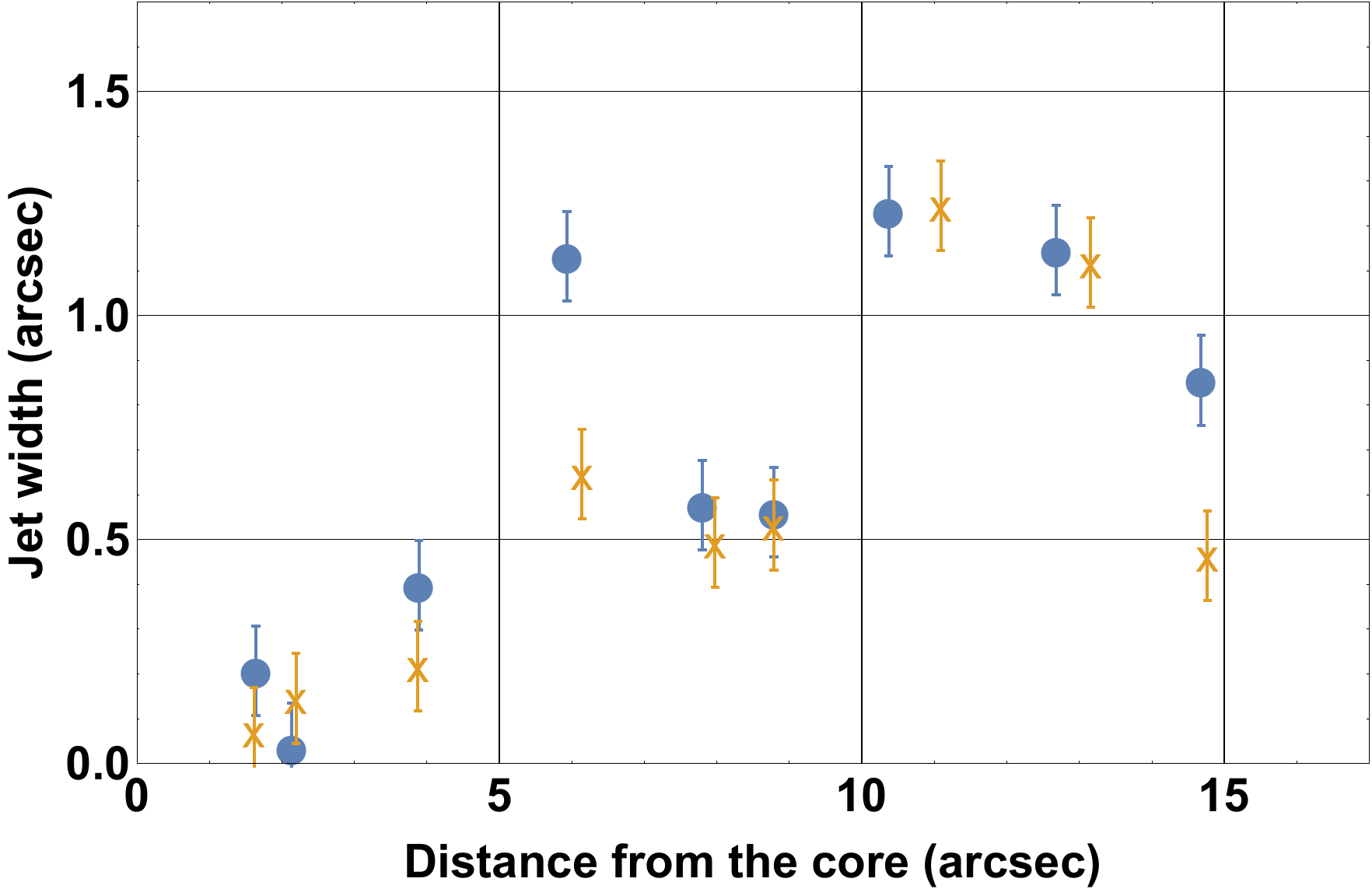}
\caption{Dependence of position angle ({\it top}) and width ({\it
bottom}) of the jet on distance from the core at 5~GHz (filled
circles) and 15~GHz (crosses). These quantities are measured from 2-d
components using
modelfit in difmap. The jet appears wider than in X-rays, but note the
different modeling technique (1-d cross-cuts) in the X-ray
case. \label{fig:jetwidth}}  
\end{figure}

\begin{figure*}[t]
\epsscale{0.9}   
\plotone{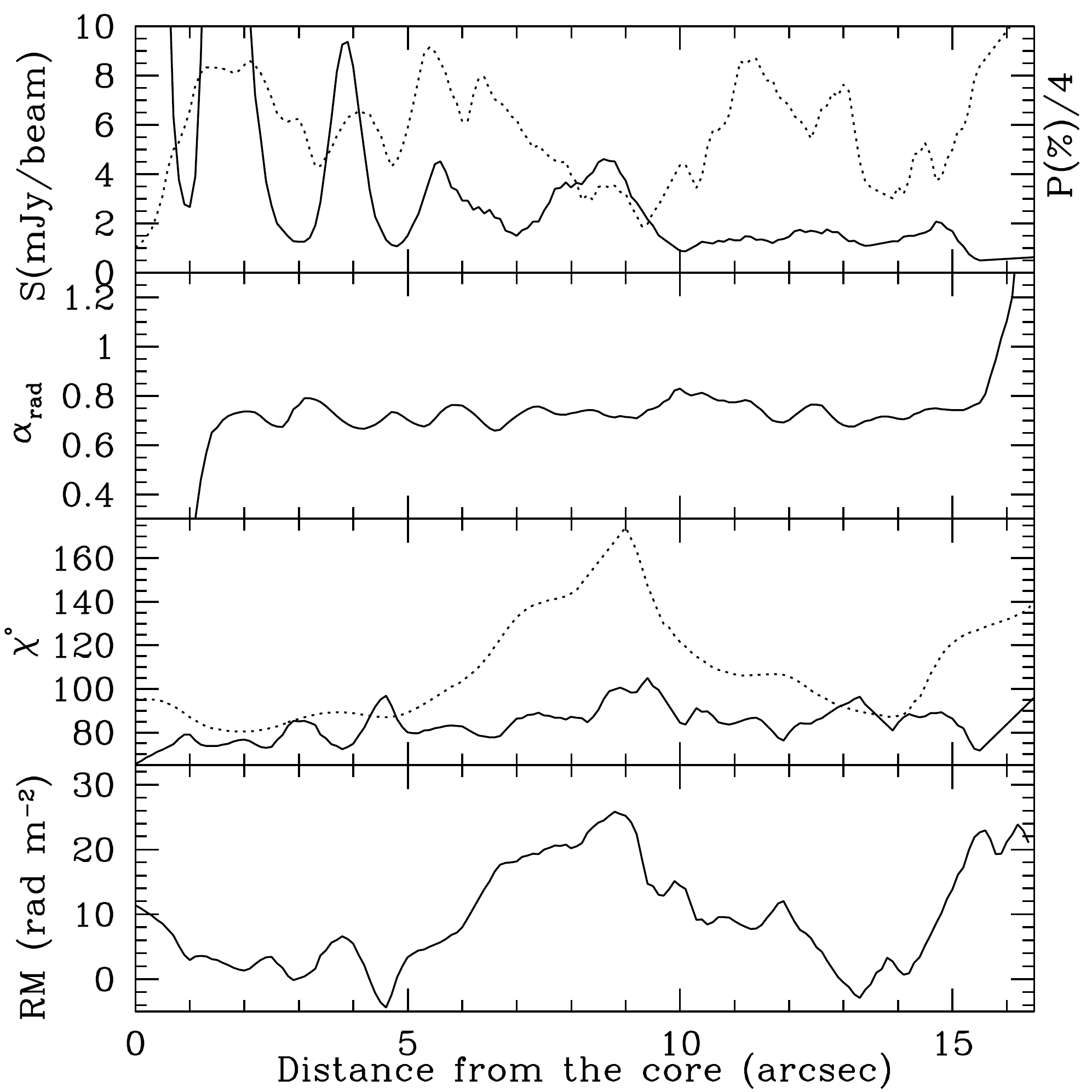}
\caption{Radio profiles ({\it from top to bottom}) of: 1. The total
  intensity (solid curve) and percent polarization (values are
  divided by a factor of 4, dotted curve) at 5~GHz; 2. The spectral index
  between 1.4 and 15~GHz; 3. The position angle of polarization
  (electric vector) at 5~GHz
  (solid curve) and 1.4~GHz (dotted curve); 4. The Faraday rotation
  measure. \label{fig:jetprofs}}
\end{figure*}

\begin{table}
\begin{center}
\caption{Radio and X-ray Spectral Results for the X-Ray Jet
  Regions\label{tab:xspec}} 
\begin{tabular}{rllrc}
\tableline\tableline
Region & $\alpha_{R }$ &$\alpha_{Rerr}$ &  $\alpha_{\rm X}$  &    $\alpha_{\rm Xerr}$ \\
    &     &  1 sigma & &  1 sigma  \\
\tableline
 S4.0  & 0.74 & $\pm$0.05  &   0.55  &$\pm$0.13    \\
 S5.3  & 0.74 & $\pm$0.09  &   0.90  & $\pm$0.22 \\    
 S6.6  & 0.78 & $\pm$0.07  &   0.87  & $\pm$0.18 \\
 S8.3  & 0.82 & $\pm$0.05  &   0.64  & $\pm$0.14 \\
 S10.0 & 0.88 & $\pm$0.14  &   0.61  & $\pm$0.20  \\
 S11.2 & 0.87 & $\pm$0.15  &   0.68  & $\pm$0.21  \\
 S12.9 & 0.78 & $\pm$0.09  &   0.93  &  $\pm$0.16 \\
 S14.6 & 0.86 & $\pm$0.09  &   0.88  &  $\pm$0.17  \\
 S15.9 & 0.83 & $\pm$0.27  &   0.81  &  $\pm$0.16  \\
 S17.7 & 0.93 & $\pm$0.47  &   0.81  &   $\pm$0.13  \\

 \tableline
\end{tabular}
\end{center}


\end{table}

\begin{table}
\begin{center}
\caption{Summary of the VLA Observations\label{tab:radobs}}
\begin{tabular}{rcllcccc}
\tableline\tableline
Frequency & Array & TOS\tablenotemark{a} & Clean Beam\tablenotemark{b} &I Map Peak&I rms&P Map Peak&P rms\\
(GHz) & & &  
&(mJy/Beam)& (mJy) &(mJy/Beam)& (mJy) \\
\tableline
1.365 & A & 4$^h$10$^m$ & 1.21$\times$1.07, 50.6 &1240& 0.012&71.9&0.063 \\
1.435 & A & 4$^h$10$^m$ & 1.16$\times$1.03, 48.4 &1210&0.028&71.1&0.075 \\
4.860 & A & 4$^h$10$^m$ & 0.47$\times$0.43, 38.5 &1164& 0.073&49.6&0.039 \\
4.860 & B &3$^h$31$^m$ & 1.09$\times$1.06, 19.2 &1171& 0.061&51.5&0.078 \\
14.940& B &5$^h$21$^m$ & 0.50$\times$0.49, 36.4 &1805& 0.096&34.0&0.088 \\
\tableline
\end{tabular}
\end{center}
\tablenotetext{a}{TOS is the effective time on source}
\tablenotetext{b}{The clean beam size gives the major$\times$minor
  axis in arcsecs and the position angle of the major axis in
  degrees.  Uniform u,v weighting was used for all maps.}

\end{table}

\subsubsection{Radio Jet\label{radiojet}}

Figure~\ref{fig:jetmaps} shows total intensity images of the jet with
linear polarization electric vectors.  The jet has knotty structure,
which we have modeled as circular Gaussian brightness distributions
using the task MODELFIT in {\it Difmap}.  For each component we have
obtained angular distance, $R$, and position angle,
$\Theta$, relative 
to the core located at RA: 13$^{\rm h}$57$^{\rm m}$04.4366$^{\rm s}$ and
DEC:+19$^\circ$19\arcmin 07\farcs372 ~(J2000), as well as FWHM size, $w$, and
flux density, $S$. The number of components required to fit the data
was determined by the best agreement between the model and data
according to $\chi^2$ values.  In general, 10 Gaussian components give
a reasonable representation of the jet morphology at all frequencies,
with reduced $\chi^2$ ranging from 1.5 to 5.
Figure~\ref{fig:jetwidth} shows the dependence of $\Theta$ and $w$ on
distance from the core.  We estimate 0\farcs1 uncertainty in position
or width, so that the position angle error is 0\farcs1/(distance in
radians). Although the jet executes wiggles close to
the core, deviations do not exceed the uncertainties, and, on average,
the jet is straight within 9\asec ~of the core, with
$\Theta$=165.0$^\circ\pm$0.5$^\circ$.  Beyond 9\asec ~it turns to the
east by 1.5$^\circ\pm$0.5$^\circ$ relative to the core, or
$\sim$4$^\circ$ with respect to the previous direction. The independence
of the width of the jet on distance indicates that the jet is well
collimated, with similar size near the core and at the end (at 17\asec
~or 130~kpc in projection from the core). However, in the transverse
direction the jet appears to undergo contractions and
expansions. These could be the result of standing shock formation in
the jet flow as it adjusts to imbalances between the jet and external
medium pressures, as seen in numerical simulations \citep{aloy03}.
Both properties (good collimation and contraction-rarefaction
structure) imply that the jet should be highly supersonic and probably
relativistic far downstream of the core.

We have constructed profiles along the jet axis of total intensity,
degree of polarization, spectral index, position angle of
polarization, and Faraday rotation measure ($RM$), as plotted in
Figure~\ref{fig:jetprofs}. Each point of a profile is the median average
within a window of half a beam size. The window slides by half of its
size for every new measurement.  There is a prominent feature in the
polarization profiles in the region from 7\asec ~to 10\asec. The
profiles at all wavelengths show an increase of total intensity, a
sharp decrease of degree of polarization, and a change of position
angle of polarization in this region. In addition, the change of
position angle of polarization depends on wavelength.  Such behavior
can result from an increase of $RM$, in this part of the jet. We have
calculated the $RM$ values using polarization maps at 1.365, 1.435,
4.86, and 14.965~GHz, convolved with the same circular beam of
1\asec$\times$1\asec ~for the polarized intensity exceeding the
3$\sigma$~rms level. At 14.965~GHz this condition applies only within
$\sim$5\asec ~from the core. The average rotation measure is low,
$\sim~5$~rad~m$^{-2}$,  being consistent with integrated measurements
and likely of Galactic origin \citep{Simard81}. However, the 
 RM increases by a factor of 2 in the
core and by a factor of 5 in the region from 7\asec ~to 10\asec ~and
at the end of the jet. The HST image (see Figure~\ref{fig:opt3pan})
contains a galaxy partially projected on the jet 6\asec-7\asec ~from
the core. The gas from the galaxy could cause the observed
polarization behavior if the galaxy lies along the line of sight to
the quasar.  On the other hand, the brightening is difficult to
explain if the galaxy is intervening rather than interacting with the
jet. In the former case, the increase in the intensity must be
intrinsic. Because of this ambiguity, we exclude the region affected
by the galaxy and also the core region within 1\farcs5 (the core has
degree of polarization  $P^{\rm core}_{\rm 5~GHz} \sim$4\%) from all further
discussion of the jet radio properties.

\begin{figure*}[t]
\includegraphics[width=0.9\columnwidth]{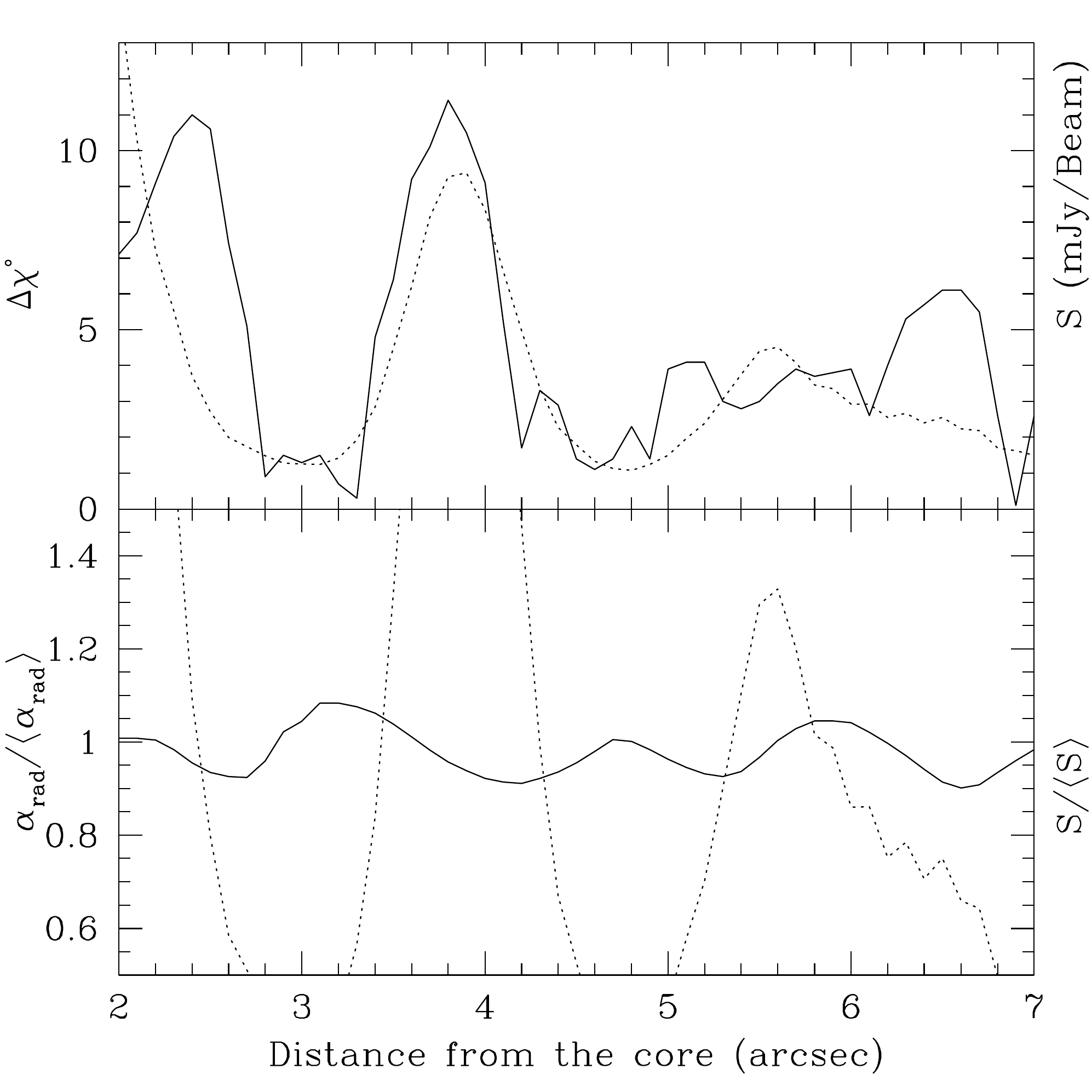}
\caption{{\it Top:} Profiles of variations of polarization position angle (solid curve) and total intensity
(dotted curve) at 5~GHz. {\it Bottom:} Normalized profiles of spectral index (solid curve), 
$\langle\alpha_{\rm r}\rangle$=0.73, and total intensity (dotted
curve), $\langle S\rangle$=3.4~mJy/Beam. \label{fig:jetdevs}} 
\end{figure*}

The total intensity along the jet varies by a factor of $\sim$50 while
the degree of polarization changes by a factor of $\sim$3. There is a
tight correlation between the total and polarized intensity in the
jet, with linear coefficient of correlation $\rho=0.94$; however, the
degree of polarization does not correlate with total intensity
($\rho=0.13$). On average, the jet is highly polarized, with $\langle
P_{5 GHz}\rangle$=(24.7$\pm$6.0)\% and polarization position angle
$\langle\chi_{5 GHz}\rangle$=(84$\pm$6)$^\circ$. These values
indicate that the magnetic field aligns with the jet direction and
that the degree of field order remains fairly uniform along the
jet. Nevertheless, comparison of the 
variations in polarization position angle (Figure~\ref{fig:jetdevs}),
defined as $\Delta\chi=\left\|\chi-\langle\chi\rangle\right\|$, with the
total intensity behavior along the jet suggests a positive correlation
between the two ($\rho=0.37$). This correlation can be explained if
the magnetic field tends to be more turbulent in bright knots than in
the underlying jet. We caution, however, that the maximum variations
of $\chi$ are only $\sim$10$^\circ$, while the uncertainties of
individual measurements of polarization position angle are $\sim$5$^\circ$.

\begin{figure*}[t]
\includegraphics[width=0.45\columnwidth]{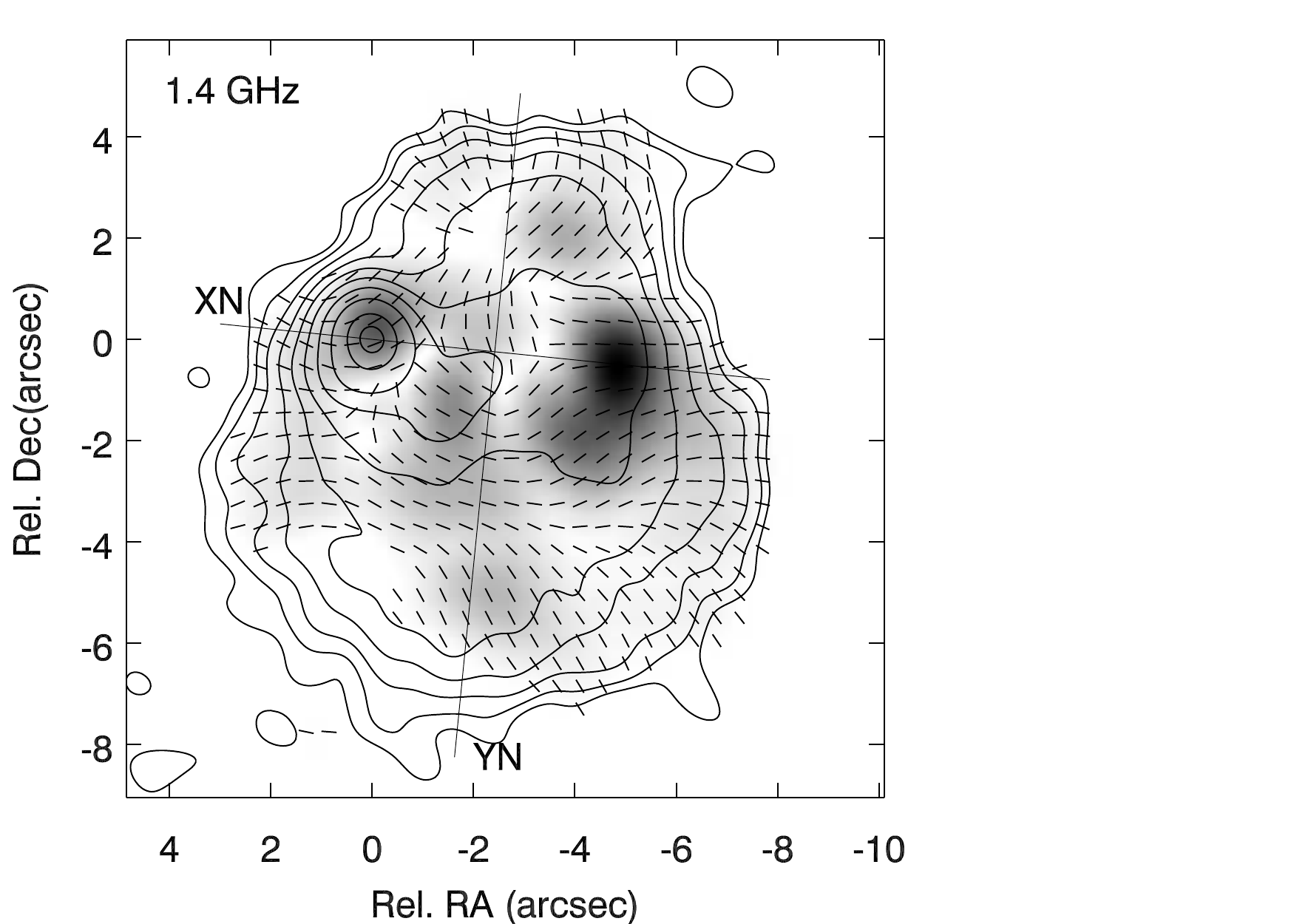}
\includegraphics[width=0.45\columnwidth]{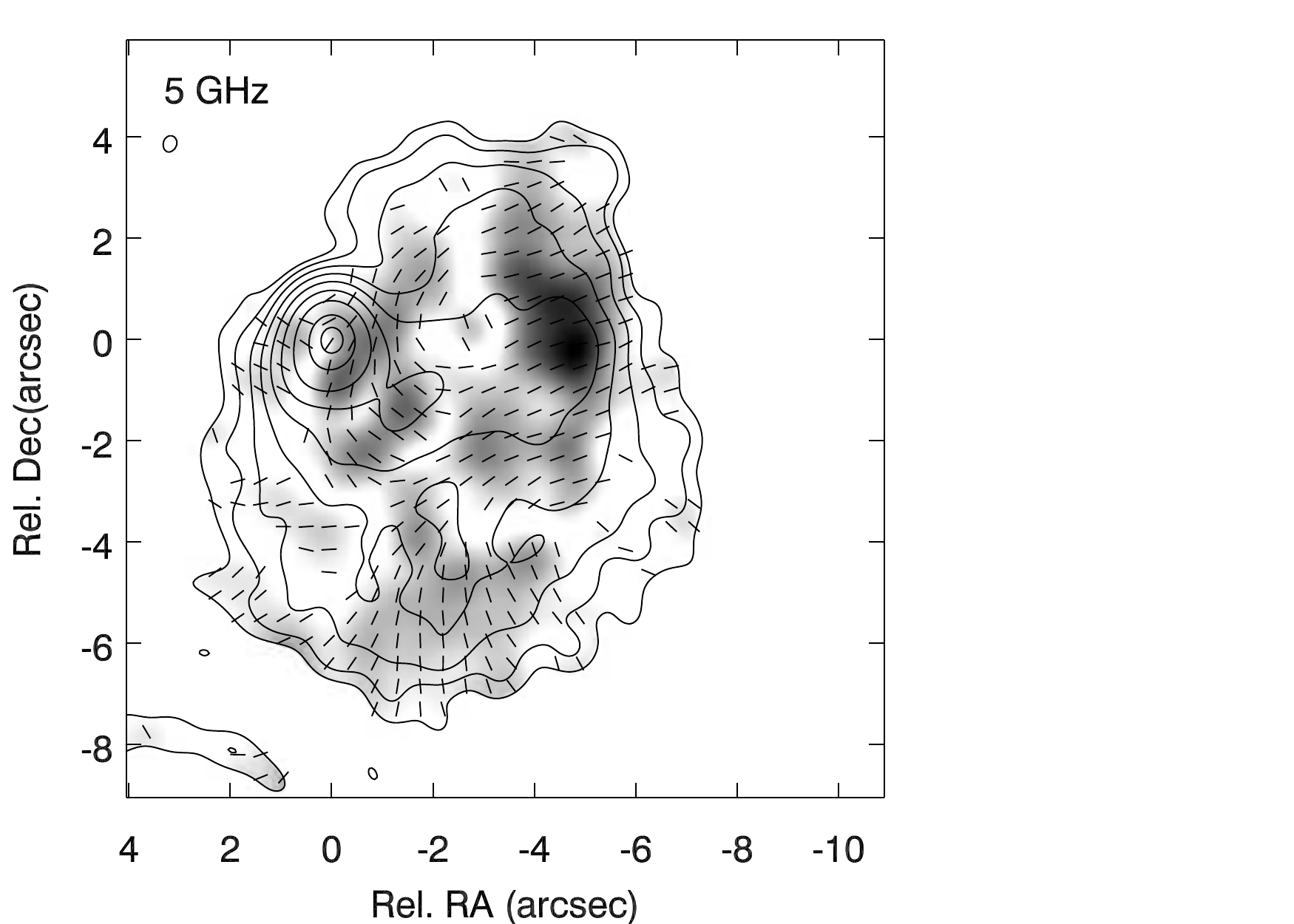}\\
\includegraphics[width=0.4\columnwidth]{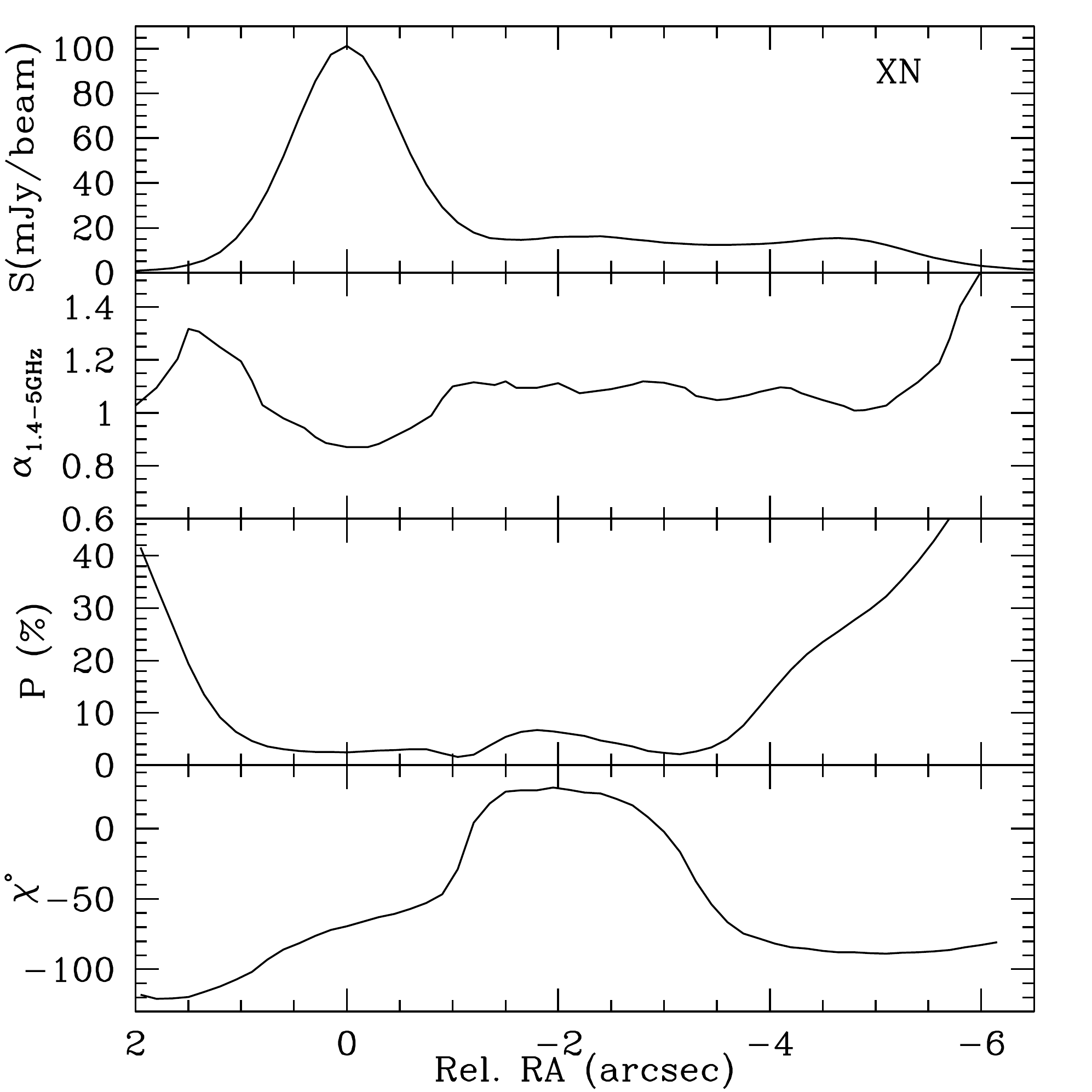}q
\includegraphics[width=0.4\columnwidth]{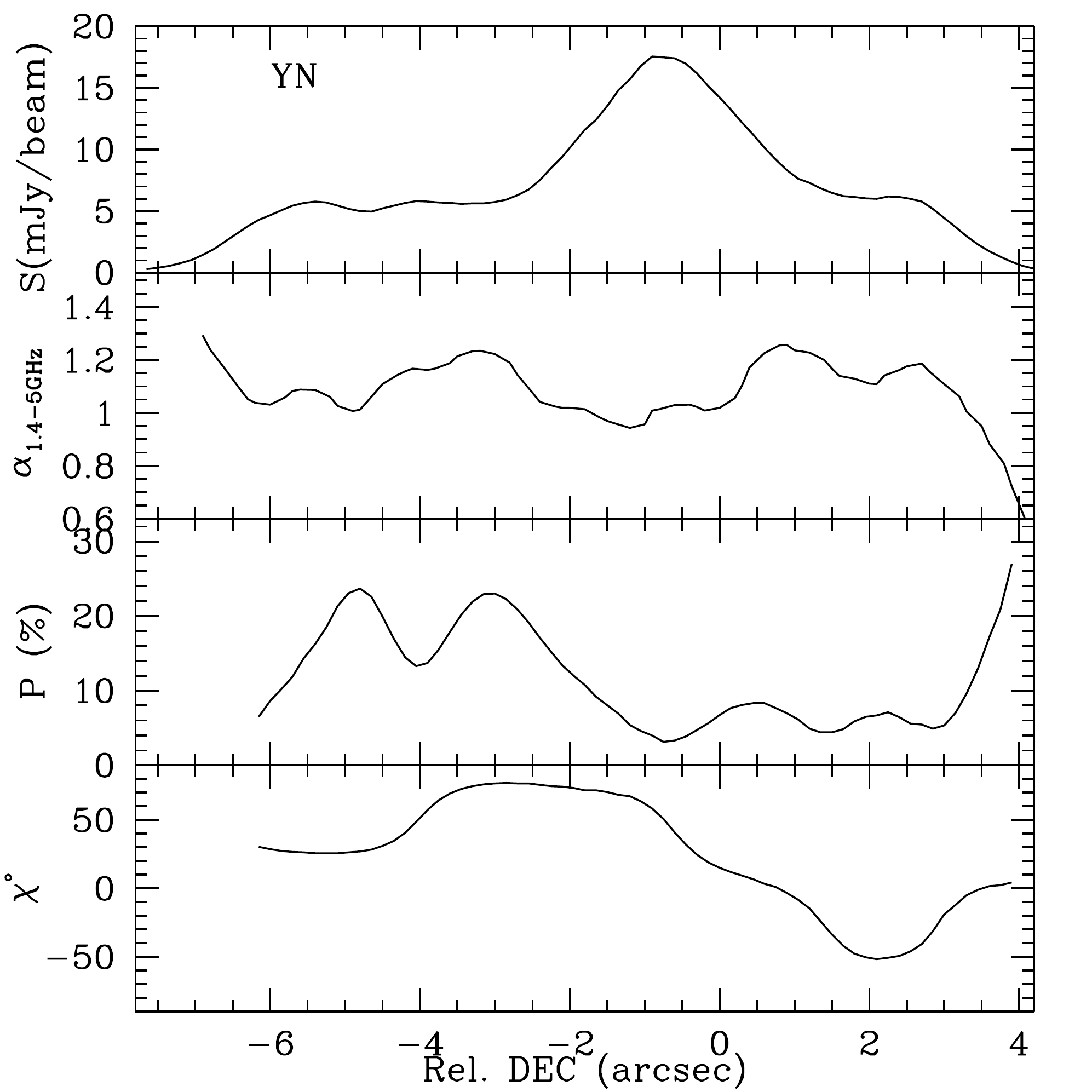}
\caption{{\it top:} Total (contours) and polarized (gray scale) intensity
  images of the northern lobe. {\it left:} at 1.4~GHz (A-array),
  $S_{\rm peak}$=101.3~mJy/beam, $S_{\rm peak}^{\rm pol}$=4.16~mJy/beam, and {\it right:}
  at 5~GHz (B-array), $S_{\rm peak}$=33.4~mJy/beam,
  $S_{\rm peak}^{\rm pol}$=1.81~mJy/beam. Beam size is according to
  Table~\ref{tab:radobs}; linear segments correspond to polarization E-vectors;
  solid lines in left panel show two axes XN and YN used for constructions of profiles in the
  lower left and right panels, respectively. {\it bottom:} Those profiles show
  (top to bottom) 1. Total intensity at 1.4~GHz; 2. Spectral index between 1.4
  and 5~GHz; 3. Degree of polarization at 1.4~GHz; 4. Position angle of
  polarization at 1.4~GHz. \label{fig:Nmaps}}
\end{figure*}

We have calculated the radio spectral index, $\alpha_{\rm R}$, using
flux densities at 1.365, 1.435, 4.86, and 14.965~GHz from images
convolved with the same beam of 1\asec$\times$1\asec\ .
We find an inverted spectrum in the core $\alpha_{\rm
  R}^{core}=-0.14\pm 0.07$, an optically thin spectrum in the jet,
$\langle\alpha_{\rm R}\rangle=0.82\pm0.06$, and a steeper spectrum
near the end of the jet (beyond 15.5\asec), ~$\alpha_{\rm R}>$1.0.
There are small variations of $\alpha_{\rm R}$ along the jet between
1\farcs5 and 15\farcs5 from 0.74 to 0.93 with the uncertainty of
individual measurements from $\pm$0.05 up to $\pm$0.5. These
variations show a possible anti-correlation with total intensity
($\rho=-0.30$, see also Figure~\ref{fig:jetdevs}), which implies a
slightly harder radio spectrum within bright knots.
Table~\ref{tab:xspec} gives the separate radio and X-ray spectral
indices and their 1$\sigma$ errors in the ten regions used for X-ray
photometry of the jet. All are consistent with the value 0.80,
considering the uncertainties.  The X-ray spectral index of the quasar
core is slightly flatter, $\alpha_{\rm x} = 0.66^{+0.07}_{-0.06}$
(Marshall et al. 2017, ApJS, submitted).

\subsubsection{Radio Lobes}

\begin{figure*}[t]
\includegraphics[width=0.45\columnwidth]{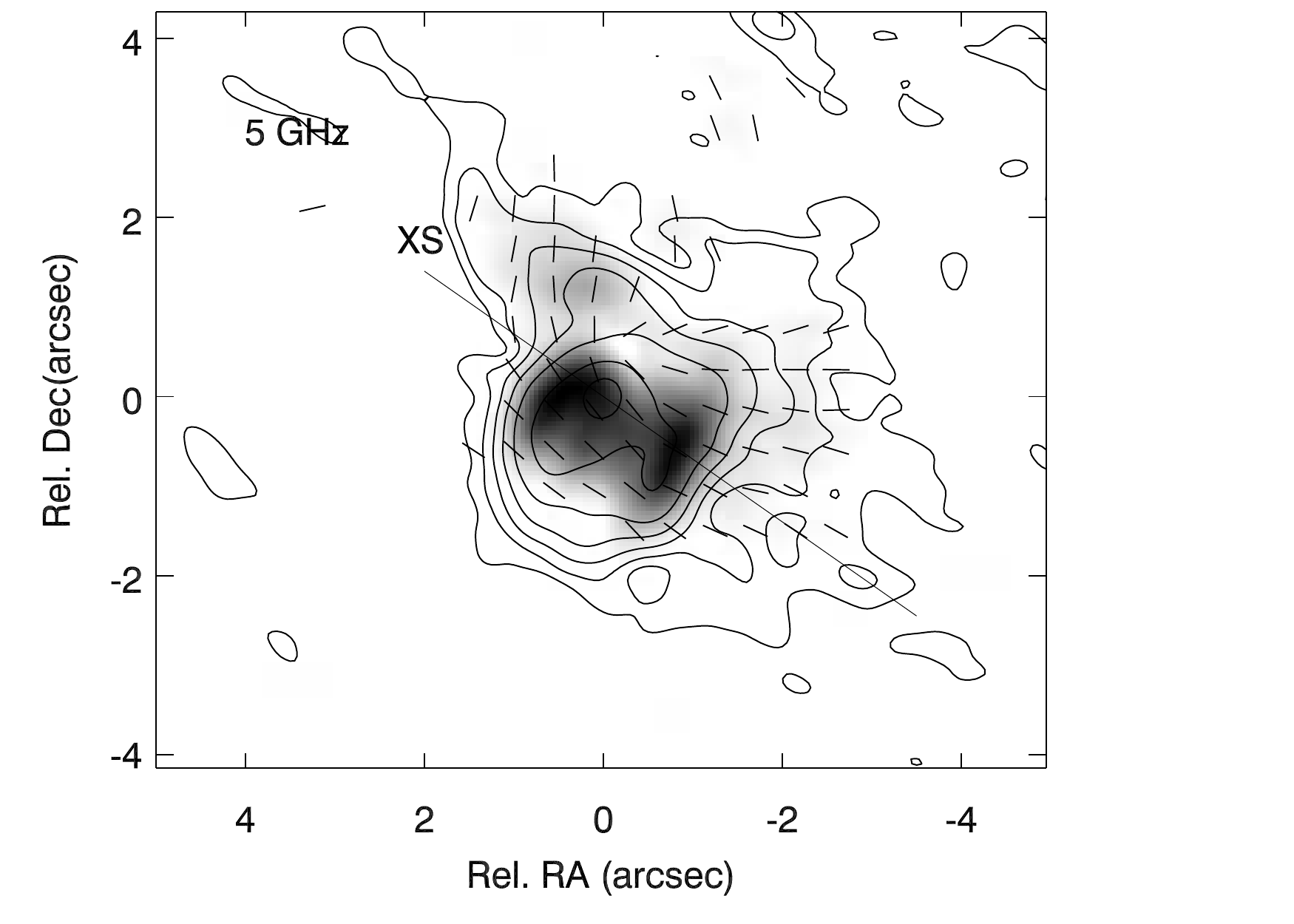}
\includegraphics[width=0.45\columnwidth]{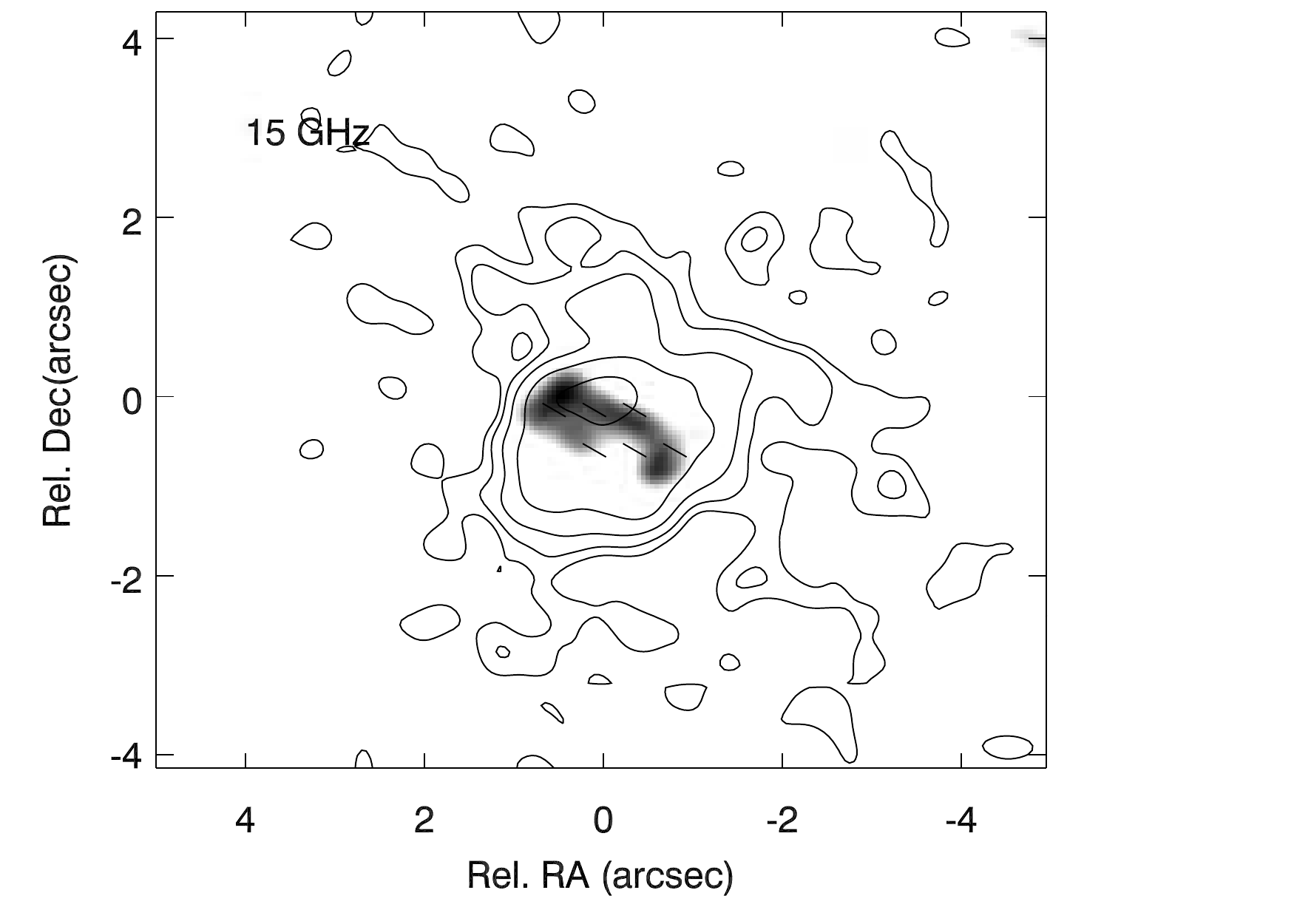}
\includegraphics[width=0.5\columnwidth]{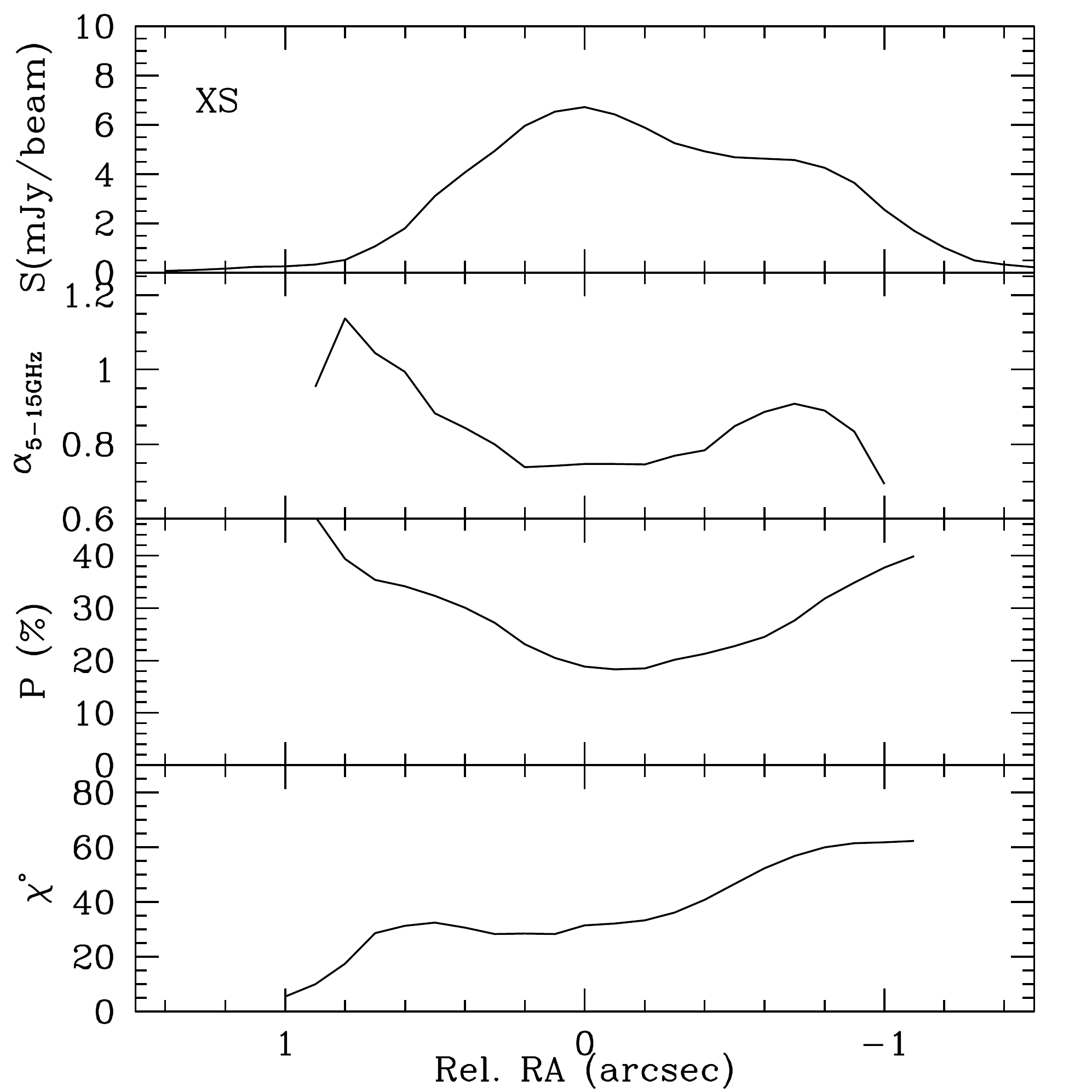}
\caption{Top: Total (contours) and polarized (gray scale) intensity
  images of the southern lobe {\it left:} at 5~GHz (A-array),
  $S_{\rm peak}$=6.76~mJy/beam, $S_{\rm peak}^{\rm pol}$=1.40~mJy/beam, and {\it
    right:} at 15~GHz (B-array), $S_{\rm peak}$=2.94~mJy/beam,
  $S_{\rm peak}^{\rm pol}$=0.89~mJy/beam. Beam size is according to Table
  \ref{tab:radobs}; linear segments correspond to polarization
  E-vectors. Bottom: Radio profiles of the southern lobe along axis XS
  in the top left panel.  From
  top to bottom; Total intensity at 5~GHz; Spectral index between 5
  and 15~GHz; Degree of polarization at 5~GHz; Position angle of
  polarization at 5~GHz. \label{fig:Smaps}}
\end{figure*}

The northern radio lobe (Figure~\ref{fig:radmap}) is located
$\sim$16\asec ~from the quasar at the position angle \\
$\Theta$~$\sim$~-~14$^\circ$, which corresponds to the direction of the
inferred counter-jet. Figure~\ref{fig:Nmaps}
shows the total and polarized intensity structure of the lobe. There
are two prominent features: a bright knot in total intensity (hot spot)
and a separate bright knot in polarized intensity (polarized spot) located
~4\asec ~to the west from the hot spot. We have constructed the total
intensity and degree of polarization profiles at 1.4~GHz along the
line connecting the knots (axis XN) and along the line perpendicular
to the juncture (axis YN) (see Figure~\ref{fig:Nmaps}), as well as
spectral index profiles using 1.4~GHz and 5~GHz maps obtained with
similar resolution beams.  The spectral index profiles show that
$\alpha_{1.4-5 GHz}$=1.1$\pm$0.1 dominates the diffuse part of the lobe
except two regions: i) the region at the western end beyond the
polarized spot with the steepest spectrum ($\alpha\sim$1.4) and ii)
the region on the northern end with the flattest spectrum ($\alpha\sim
0.7$). The hot spot also has a fairly hard spectrum, $\alpha\sim
0.8$. The degree of polarization in the hot spot is low, $P\sim2\%$,
perhaps due to rotation of the magnetic field direction within the
beam as seen from the $\chi$ profile. The northern edge has an
increase of degree of polarization up to 30\% and position angle of
polarization perpendicular to the boundary, i.e., the magnetic field
ordered along the boundary (assuming a small $RM$).  These conditions,
along with hardening of the spectrum imply a shock formation on the
northern end.  It is difficult to understand the nature of the
polarized spot that has similar surface brightness to the neighboring
region, a steep spectrum ($\alpha \sim 1.1$), high polarization
$\sim$20-30\%, and fairly uniform magnetic field along the YN axis.
In general, the magnetic field in the lobe has patchy structure, with
the size of a patch $\sim$2\asec$\times$2\asec ~and uniform magnetic
field within a patch, with changing direction from one patch to
another.

The diffuse part of the southern lobe is most likely located between
the end of the jet and the southern hot spot, with too low a radio surface
brightness to be seen in our high resolution maps,
(e.g. Figure~\ref{fig:radmap}).  The hot spot is located 28\asec ~from the
core at $\Theta\sim$170$^\circ$, shifted by $\sim$5$^\circ$ from the
jet direction. The high resolution maps (Figure~\ref{fig:Smaps}) show
that the hot spot has a double structure with a separation between
peaks $\sim$0.7\asec. We have constructed profiles along the line
crossing the peaks, (axis XS in Figure~\ref{fig:Smaps}), similar those
obtained for the northern lobe; the profiles are presented in the
bottom section of 
Figure~\ref{fig:Smaps}. The spectral index ($\alpha_{5-15 GHz}$) of
the region is similar to that of the jet. The magnetic field is rather
uniform with the direction perpendicular to XS axis, the polarized
intensity increases in the peaks, and the whole structure has an
almost constant degree of polarization $\sim$20\%.

\subsubsection{VLBA 2cm Survey Data}

\begin{figure*}[t]
\includegraphics[width=4.in]{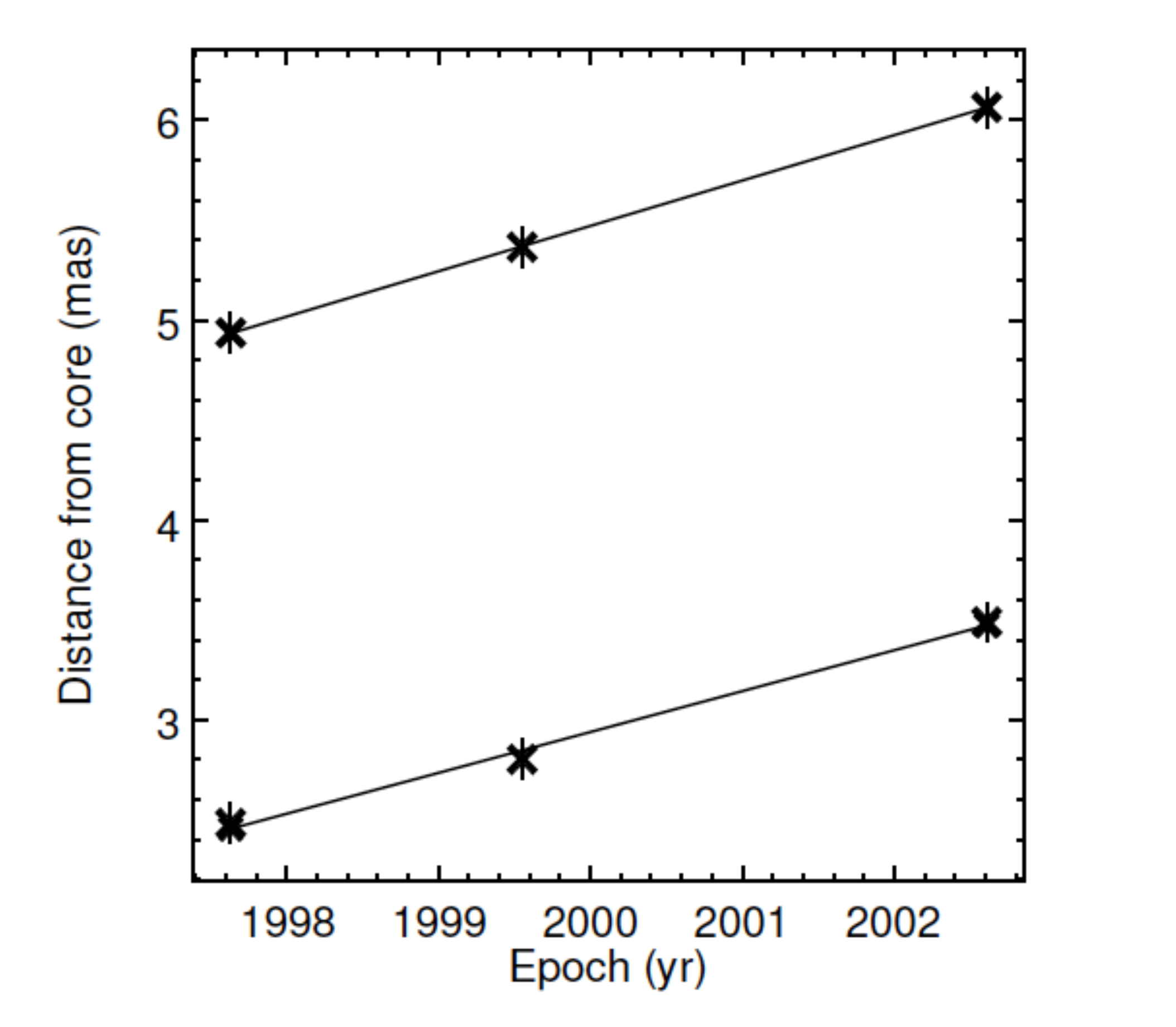}
\caption{Proper motions of two features in the VLBA jet.  The observed
  motions correspond to velocities of 8.1$\pm$1.2c (inner feature, bottom) and
  9.4$\pm$1.2c (outer feature, top).\label{fig:vlba}}
\end{figure*}

Three epochs (1997.63, 1999.55 and 2002.61) of 15 GHz VLBA
observations 
were obtained as part of the VLBA 2 cm survey \citep{kell04,zens02}.
Since no published VLBI proper motions were available in the
literature, we obtained the calibrated ($u,v$) data from the MOJAVE
website\footnote{http://www.physics.purdue.edu/MOJAVE/}
\citep{Lister09} and modeled them with circular Gaussians using {\it Difmap}
to fit knot positions. Two distinct knots are found with average
separations of $\sim$3 mas and $\sim$5.5 mas from the core during
these epochs.  These components are well fitted with proper motions of
0.20 mas/yr and 0.23 mas/yr, at average position angles of 145\degr and
142\degr, respectively (c.f. PA=165\degr\ for the kpc scale jet); see
Figure~\ref{fig:vlba}. These translate to apparent motions of
$(8.1 \pm 1.2)c$ and $(9.4 \pm 1.2)c$, where the uncertainties assumed 0.1
mas errors in determinations of the knot positions. \citet{Marshall17}
use an additional VLBA measurement in 2003.01 to estimate velocities
8.68$\pm$0.4 c and 9.84$\pm$0.7 c, respectively, for the inner and outer
knots. 

The apparent superluminal proper motions require the pc-scale jet to
be aligned at  $< 14\degr$ (3 mas knot) and $< 12\degr$ (5.5 mas knot)
to our line-of-sight. The observed difference in the projected
position angles of the outermost VLBI-scale knot and the  kpc-scale
jet of  $23\degr$ is likely caused by a smaller intrinsic bend in the
jet amplified by projection. For small observed misalignments, the
intrinsic bend in the jet is probably smaller than the angle to the
line-of-sight \citep[see][]{Conway93,Marshall11,Singal16}. In this
case, we take an 
intrinsic bend of  $\sim 23\degr$ sin ($12 \degr$)  $\sim 5\degr$ in the
jet to estimate that the kpc-scale jet is likely aligned at $\leq
12\degr$ to our line-of-sight, consistent with estimates to be
presented in Section \ref{sec:disc}.

\subsection{Comparison of X-ray and Radio Profiles}

\begin{figure}
\epsscale{.80}
\plotone{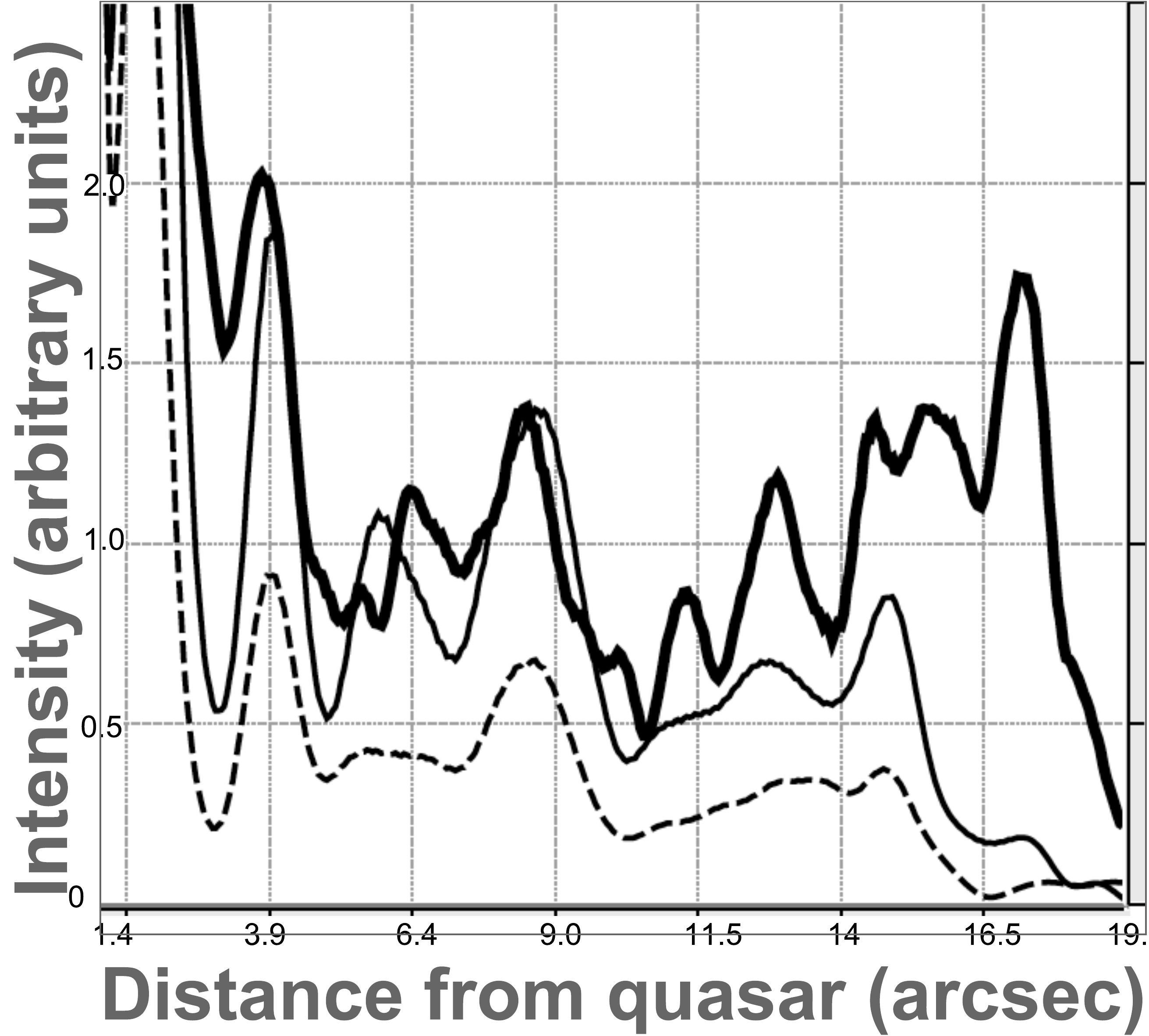}
\caption{X-ray and radio profiles down the jet.  The profile region
  is 2$^{\prime\prime}$ wide and 18$^{\prime\prime}$ long. The
  horizontal axis starts 0\farcs87 from the quasar and the region S2.1
  is off the top scale. The heavy line
  is the X-ray (0.4-6 keV), the lighter line is for 5 GHz, and the
  dashed line is 15 GHz.  The \chan\ image was smoothed with a
  Gaussian of FWHM=0.5$^{\prime\prime}$, producing an effective
  resolution of about 0.85$^{\prime\prime}$.  The radio maps had clean
  beams of 0.5$^{\prime\prime}$, but were smoothed with Gaussians
  suitable to produce a beam size equivalent to the \chan\ resolution.
  The X-ray map was fractionally binned in order to match the radio
  pixel size of 0.05$^{\prime\prime}$.\label{fig:profile}}
\end{figure}

To compare radio and X-ray structures we constructed profiles of the
X-ray, 5 GHz, and 15 GHz emission of the main jet (see
Figure~\ref{fig:profile}). 
We first filtered the merged event files for the energy band 0.4-6
keV.  Next we binned the data so as to match the pixel size of the
radio maps.  Since our event file had previously been registered so as
to align the X-ray and radio nuclear positions, we employed a
projection region in ds9 \citep{Joye03} based on WCS coordinates: 2$^{\prime\prime}$
wide and 18$^{\prime\prime}$ long.  We then scaled the X-ray profile
by a factor of 0.01 so that both radio and X-ray curves could be
easily compared.  Before performing the profiles, we smoothed the
X-ray data with a Gaussian of FWHM=0.5$^{\prime\prime}$ in order to
minimize statistical fluctuations.  For an intrinsic beam size of
0\farcs75, the resulting map had an effective resolution of $\approx$
0\farcs85. We then applied an appropriate Gaussian smoothing to the
two radio maps which originally had clean beams of 0\farcs5.

Within a factor of 2, the X-ray and radio profile shapes are similar,
but with differences larger than those between the 5 GHz and 15 GHz
profiles.  For the most part, the brighter X-ray enhancements can be
associated with corresponding radio knots, but not necessarily at
identical positions.  However, towards the end
of the main jet, ($>15\arcsec\ $ from the quasar), there is a marked
departure: the X-ray intensities become larger going downstream
whereas the radio fades away by a relative factor of 10.  The overall
comparison is in stark contrast to the 3C 273 jet which is X-ray bright
at the upstream end, and then drops by a factor of 100 relative to the
radio jet that continuously brightens going away from the quasar.  For
three prominent enhancements, there appears to be a small offset (of
order 0\farcs2, or 1.4 kpc in the plane of the sky) of the peak
brightness in the sense that the X-ray peaks upstream of the radio, as
commonly seen in FR I jets \citep{Hardcastle01,Dulwich07}, and also in
FR II jets, e.g., 3C353 \citep{Kataoka08} and quasars, e.g.,
PKS1127-145 \citep{Siemiginowska07}.  These knots (N to S) are located
at distances 3\farcs9, 8\farcs5, and 14\farcs5 from the quasar.  There
are also jet segments for which the X-ray intensities do not track the
radio.  The most obvious such is the radio peak 5\farcs6 from the
nucleus, with the X-ray peak downstream at 6\farcs4 in the figure.
This is near the region where a galaxy overlaps the jet, and we note
that the 5 GHz and 15 GHz profiles are also dissimilar there.
 
\section{Summary}
We summarize the key features of the data presented above:
\begin{itemize}
  \item X-rays trace the radio jet along a projected length of at
    least 140 kpc in the plane of the sky.
    \item The jet is very nearly straight out to 18\arcsec\ from
      the quasar, with an apparent projected bend of  about 4\degr\
      past 10\arcsec\ . The intrinsic bend is probably smaller, due to
      the small angle of the jet to our line of sight. 
      \item The radio and X-ray profile shapes track within a factor
        of 2 along the straight jet from 4\arcsec\ to 14\arcsec\ but
        cases of X-ray peaks upstream and downstream of radio peaks
        both occur. 
        \item The jet appears broader in the radio than the X-ray in
          the region 5\arcsec\ to 15\arcsec\ from the core. 
        \item The jet is at $<$ 12\degr\ to the line of sight.
    \item Radio and X-ray spectra are consistent with an average energy index
      0.80$\pm$0.1. 
    \item The jet likely remains relativistic far downstream of the core,
as inferred from the collimation and the contraction-rarefraction structure.
      \item The magnetic field aligns with the jet and remains fairly
        uniform along the jet.
    \item Correlation between the total intensity and the dispersion
      in polarization
      position angle suggests that the magnetic field tends to be more
      turbulent in the radio knots.
    \item There are double hot-spots in  the  southern radio
      lobe, and separate hot spots in intensity and in polarization in
      the northern radio lobe. Polarization  and spectral hardening
        indicates shock formation at the edge of the northern radio lobe.
\end{itemize}

\section{Discussion and Conclusions\label{sec:disc}} 

\begin{figure}
\begin{center}
\includegraphics
[width=5.5in]{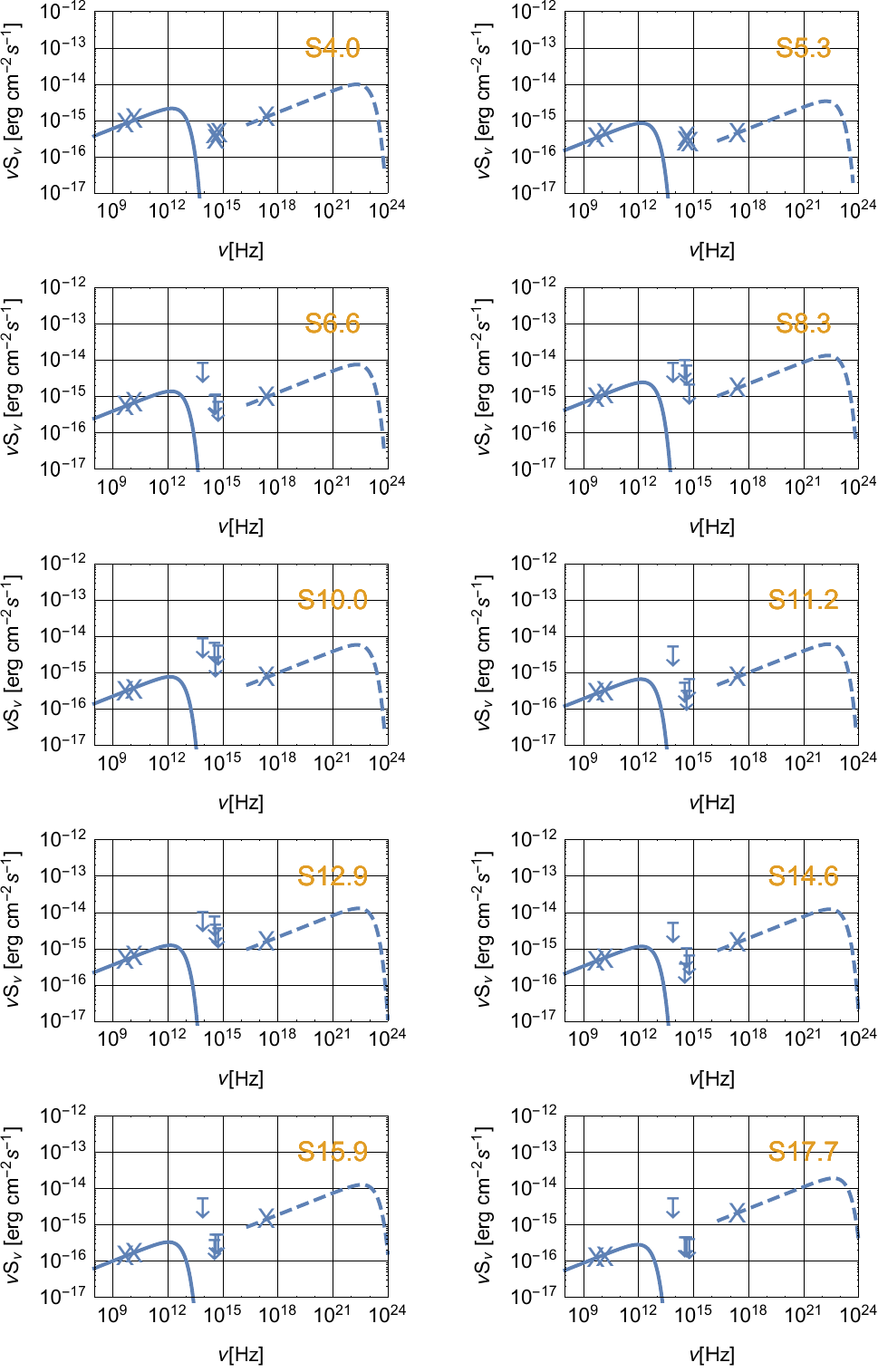}
\end{center}
\caption{Inverse Compton/CMB models fit to measured radio and X-ray
data, with an extrapolation giving predicted
  $\gamma$-ray fluxes. The models assume a power law distribution of
  electrons with spectral index 2.6 between $\gamma_{\rm min}$=30, and
  a cutoff at $\gamma_{\rm max}$=10$^5$, for the entire
  jet. Solid lines model the radio synchrotron spectra, and dashed
  lines are the iC/CMB spectra from that same electron population. \label{fig:SEDs}}
\end{figure}

We apply the iC/CMB model described by \citet{Tavecchio00} and
\citet{Celotti01} to derive the intrinsic physical conditions of the
jet. That model assumes a minimum total energy in magnetic field and
relativistic particles, and requires that the jet be in relativistic
motion with bulk Lorentz factor $\Gamma$ and be beamed at an angle
$\theta$ to our direction with a Doppler factor $\delta=(\Gamma
(1-\beta\, \cos\theta))^{-1}$. Many other assumptions are made,
including that the particles and magnetic field uniformly fill the
volume, that the relativistic electron  spectrum
dN/d$\gamma$=K$\gamma^{-m}$ gives the
observed 1.4 to 15 GHz radio emission, that the charge balance is
provided by protons that have equal relativistic energy as the
electrons, and that the angle of the jet to our line of sight,
$\theta$, takes on its maximum value for a fixed $\delta$, namely
$\arcsin(1/\delta)$, so that $\delta=\Gamma$. These are the same
assumptions made in \citet{Schwartz06}, where the sensitivity to those
assumed parameters was also calculated.  The one difference here is
that we calculate the energy in relativistic electrons by integrating
from an assumed $\gamma_{\rm min}$ of the spectrum
\citep{Worrall09,Schwartz10} instead of assuming a minimum observed
frequency in the radio synchrotron spectrum as originally formulated
by \citet{Pacholczyk70}. We choose $\gamma_{\rm min}$=30, consistent with
the result of $\gamma_{\rm min} <80 $ for PKS 0637-752 \citep{Mueller09}.
We take the volume of the regions to be cylinders of the lengths shown
in Figure~\ref{fig:regs} and Table~\ref{tab:regions}, and diameter
assumed to be 0\farcs52~=~4~kpc.

Figure~\ref{fig:SEDs} shows the measured radio, optical, and X-ray
fluxes from the 10 jet regions with detectable X-ray emission. The dashed
lines model the X-ray emission as inverse Compton scattering of the cosmic
microwave background, from the electrons giving rise to the
synchrotron spectra shown as solid lines.  We have assumed a uniform
electron spectrum with 
index m= 2$\alpha$+1=2.6, giving a mean radiation spectrum with
$\alpha$=0.80. To avoid \emph{Fermi} upper limits to GeV gamma rays
\citep[][presented $\nu f_{\nu}< 4.5 \times 10^{-14}$ at $4.6 \times
10^{22}$ Hz]{Breiding17} requires a sharp cutoff to the relativistic electrons
above $\gamma$=10$^5$. Where optical emission is detected, from S4.0 and
S5.3, it prohibits an extrapolation of the radio synchrotron spectrum
to the X-ray region \citep{Sambruna04}, as do upper limits to optical
emission in regions S6.6, S8.3, S11.2, S14.6, S15.9, and S17.7. In the other
regions the optical limits are too high to rule out such an
extrapolation; however the radio spectrum does not directly connect to
the X-rays but would over-produce the 1 keV flux density unless
cut-off at a lower frequency; e.g., in regions S10.0 and
S12.9.

In the iC/CMB model just sketched, 
  we could invoke a higher
cutoff to the electron spectrum to try to reproduce the optical
emission. A value of $\gamma \approx$ 10$^{5.8}$ would result in the
tail of the synchrotron spectrum passing close to the optical data in
regions S4.0 and S5.3.  The spectrum in the optical region would not
be well matched, and in the absence of polarization data we do not
distinguish whether such an extension or whether some additional
mechanism produces the compact optical knots.  The remainder of the
jet would still require an upper cutoff around 10$^{5}$ to avoid
exceeding the {\it Fermi} upper limits \citep{Breiding17} by the summed
iC/CMB from the entire jet.

As an alternative to the iC/CMB model, consider  whether an
additional population of electrons produces the X-ray jet via
synchrotron radiation.  In the jet rest frame, the magnetic field
energy density equals that of the CMB when
B$^2$/(8$\pi$)~=~aT$_{0}^4\,(1+z)^4\,\Gamma^2$.  At the redshift 0.72
of 4C+19.44, the magnetic field would have to be greater than 10
$\mu$Gauss, and the jet speed  would
have to be less than 0.4$c$, to exceed the CMB energy density, as
required for particles to emit primarily by synchrotron radiation
rather than inverse Compton. In such a magnetic field, 
electrons emitting 1 keV synchrotron
radiation would have $\gamma >$7$\times$10$^7$, where we use the
delta-function approximation that the particles emit at a frequency
$\gamma^2$ times the gyro frequency. 
  Since the synchrotron
frequency depends on $\gamma^2$B, while the lifetime is inversely
proportional to $\gamma$B$^2$, those electrons would have a lifetime
less than 2300 years, or a range of less than 700 pc which projects to
$<$0\farcs1 in the plane of the sky.  
By contrast, the population of  15 GHz emitting
electrons would have $\gamma \approx$10$^4$ and a range of 37 kpc
(5\arcsec\ ) in an equipartition field of $\approx$~35~$\mu$G, where
the field strength is chosen to exceed the CMB energy density even if
the jet Lorentz factor were as large as $\Gamma$=4. 
In such a scenario
it seems difficult to explain why the ratio of X-ray to radio
emission remains even within a factor of 2, over a projected distance of
115 kpc, as shown in Figure~\ref{fig:profile}. In a synchrotron model, the
X-rays are from electrons accelerated at essentially every point in
Figure~\ref{fig:profile}.  Some feedback mechanism must operate to
coordinate the separate spectra of GHz- and X-ray- emitting electrons.   The
present data does not exclude such a model, subject to the constraints
outlined above.

\begin{figure*}[t]
\includegraphics[width=0.45\columnwidth]{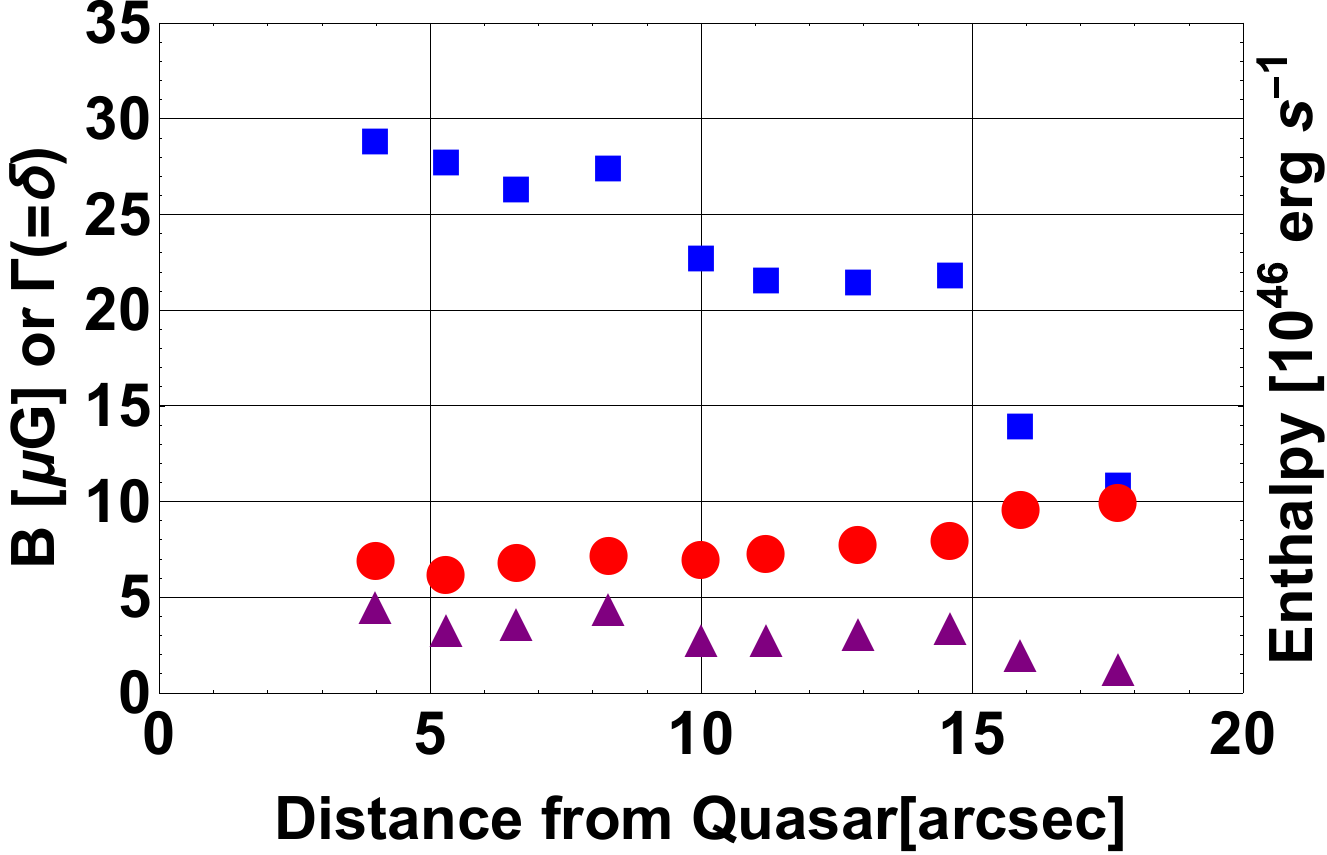}
\includegraphics[width=0.45\columnwidth]{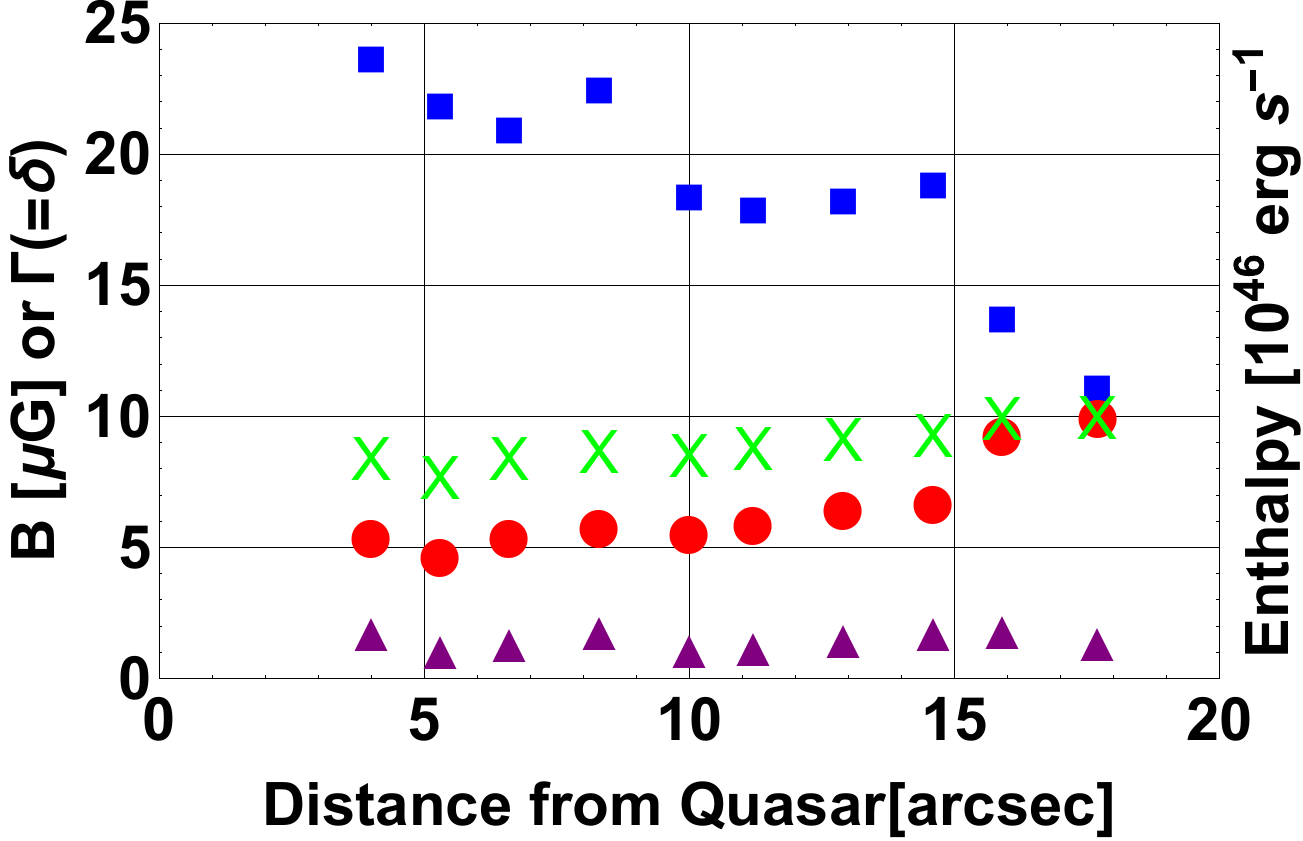}
\caption{\label{fig:bdel}Left Panel: Change of the magnetic field
  (blue squares), Lorentz factor (red dots) and enthalpy flux (purple
  triangles) as a function of distance along the jet. We assume the
  Doppler factor equals the Lorentz factor.  Right Panel: Change of
  the magnetic field (blue squares), Lorentz factor (red dots),
  Doppler factor (green crosses), and enthalpy flux (purple
  triangles) as a function of distance along the jet. For this
  calculation we assume the jet is at the constant angle
  $\theta$~=~5\fdeg8 
  to our line of sight.  Both figures assume protons provide the
  charge neutrality.}
\end{figure*}

The iC/CMB model values for magnetic field strength, B, and Doppler
factor, $\delta$, are presented in Figure~\ref{fig:bdel}. These values
are of the same order as found in other one-sided kpc X-ray jets
\citep[e.g., ][]{Sambruna02,Marshall05,Schwartz06}. Fixing the
spectral indices at the value $\alpha$=0.80 consistent with the data,
the jet shows relatively constant structure, especially from 4\arcsec\
to 15\arcsec\ from the quasar. In the left hand panel we use the
assumption that $\Gamma = \delta$ and calculate mean values $<$B$>$~=
~22 $\mu$G and $<$$\delta$$>$=7.7. The mean number density of the
minimum energy electron population would be 16.5$\times$10$^{-8}\
$cm$^{-3}$, and the angle $<$$\theta$$>$=7.6\degr\ . The deprojected
distance 10\arcsec\ from the quasar would be 580 kpc.  The kinetic
power, (enthalpy flux), is calculated assuming the charge balance is
provided by protons, which have total relativistic energy equal to
that in the electrons.  If only positrons neutralize the charge then
the kinetic flux would be about 6 times less, while the magnetic field
strength values would decrease by about 30\%.  Uncertainties in the
individual quantities are 3\% to 10\% due to photon statistics, so
systematic differences from the assumptions of isotropy of the
particles and field, of uniform volume filling factor, of low energy
electron cutoffs and that $\Gamma = \delta$, will dominate. In
particular, past 15\arcsec\ from the quasar, we calculate $\delta
\approx$ 10, implying that the jet angle is moving closer to our line
of sight, from a maximum of 9\fdeg1 to 5\fdeg7, if $\Gamma = \delta$.
But this contradicts our empirically based hypothesis that the jet is
at a constant angle to our line of sight.  If we assume instead that
the entire jet is at the minimum angle 5\fdeg7 that results from the
$\Gamma = \delta$ assumption, we calculate the run of parameters shown
in the right hand panel of Figure~\ref{fig:bdel}.  The trend of
magnetic field decreasing along the jet is still seen. Past 15\arcsec\
the magnetic field decrease, and the concomitant decrease of the
number of relativistic particles according to the minimum energy
assumption, compensates for the bulk Lorentz factor increase to
maintain a constant enthalpy flux at about 1$\times$10$^{46} $ erg
s$^{-1}$.  This could be caused by time dependent differences in the
injected structure of the jet, but at constant power. Alternately , the
divergence of the radio and X-ray jet profiles past 15\arcsec\ may
indicate a breakdown of assumptions of uniformity along the jet, or
even that the iC/CMB model does not explain all the X-ray emission in
this region. 

We have interpreted the present results in terms of the iC/CMB model
in order to estimate physical quantities in the jet. At redshifts
greater than 2.5 this must be the dominant mechanism, unless the
magnetic field strength is greater than 90 $\mu$G or the relativistic
jet speed is less than $\beta$=0.9$c$. However, at lower redshifts the
mechanism is still not certain, as has been discussed.  An alternate 
interpretation, discussed by many authors, is to produce the X-ray and
possibly optical emission by a second, high-energy electron
population, as proposed e.g., for 3C 273 by \citet{Jester06}. A
measurement of the spectral slope of the optical knots, extended into
the infrared (e.g., by JWST), and especially measurement of the optical
polarization, could indicate whether they are the extension of the
electron population producing the GHz radio emission, or due to a
distinct electron population as in 3C 273 \citep{Jester06}.
Significant improvement of the X-ray data for the 4C+19.44 jet would
require Ms \chan\ observations, which may be prohibitively expensive.
More high quality multi-band data of individual jets, as well as
larger samples of jets, are required to study the radiation and
acceleration processes in general.  Observations of high redshift X-ray jets,
where we know the emission mechanism must be iC/CMB, are particularly
needed. \chan\ is the only X-ray observatory in at least the next
twenty years which can make the required arcsec scale, high contrast
observations.

\acknowledgments 

This research has made use of the NASA/IPAC Extragalactic Database
(NED) which is operated by the Jet Propulsion Laboratory, California
Institute of Technology, under contract with the National Aeronautics
and Space Administration. This research has made use of SAOImage DS9,
developed by Smithsonian Astrophysical Observatory. We also used data
from the MOJAVE database that is maintained by the MOJAVE team.  
  We thank Anna Barnacka and the anonymous referee for comments, and
  Peter Breiding for discussion of the implications of the {\it Fermi}
  data. This work was partially supported by NASA grants GO6-7111A
(\chan\ ) and GO-10762.01-A (HST), and by NASA contract NAS8-03060 to
the \chan\ X-ray Center (D.A.S., N.L., A. S.).  Work by C.C.C. at NRL
is supported in part by NASA DPR S-15633-Y. F.M. gratefully
acknowledges the financial support of the Programma Giovani
Ricercatori - Rita Levi Montalcini - Rientro dei Cervelli (2012)
awarded by the Italian Ministry of Education, Universities and
Research (MIUR).   {\L}.S. was supported by Polish NSC grant
UMO-2016/22/E/ST9/00061

{\it Facilities: VLA, Spitzer (IRAC), HST (WFC-ACS), CXO (ACIS)} 

\clearpage 

\end{document}